\documentclass[12pt, a4paper]{article}
\usepackage{graphicx}

\usepackage{amsmath,amsfonts,latexsym}
\usepackage{graphicx}
\begin{document}

\bibliographystyle{unsrt}
\vspace{-2.5cm}

\title {Initial-value problem for coupled Boussinesq equations and \\a  hierarchy of Ostrovsky equations\thanks{Accepted for publication in {\bf Wave Motion (2011) doi: 10.1016/j.wavemoti.2011.04.003}}}
\author {K.R. Khusnutdinova\thanks{Corresponding author. Tel: +44 (0)1509 228202. Fax: +44 (0)1509 223969.},
% {\it E-mail address:} K.Khusnutdinova@lboro.ac.uk},  
K.R. Moore }
\date{}
\maketitle
\vspace{-5ex}
\begin{center}
Department of Mathematical Sciences,\\
Loughborough University,
Loughborough LE11 3TU, UK\\[2ex]
K.Khusnutdinova@lboro.ac.uk\\
K.R.Moore@lboro.ac.uk

\end{center}

\abstract{We consider the initial-value problem for a system of coupled Boussinesq equations on the infinite line for localised or sufficiently rapidly decaying initial data, generating sufficiently rapidly decaying right- and left-propagating waves. We study the dynamics of weakly nonlinear waves, and using asymptotic multiple-scales expansions and averaging with respect to the fast time, we obtain a hierarchy of asymptotically exact coupled and uncoupled Ostrovsky equations for unidirectional waves.  We then construct a weakly nonlinear solution of the initial-value problem in terms of solutions of the derived Ostrovsky equations within the accuracy of the governing equations, and show that there are no secular terms. When coupling parameters are equal to zero, our results yield a weakly nonlinear solution of the initial-value problem for the Boussinesq equation in terms of solutions of the initial-value problems for  two Korteweg-de Vries equations, integrable by the Inverse Scattering Transform.  We also perform relevant numerical simulations of the original unapproximated system of Boussinesq equations to illustrate the difference in the behaviour of its solutions for different asymptotic regimes.}
\bigskip

%{\bf PACs:} 05.45.Yv,  43.25.+y, 62.30.+d% PACS, the Physics and Astronomy


\section{Introduction}

The Ostrovsky equation
$$
(\eta_t + \nu \eta \eta_x + \gamma \eta_{xxx})_x = \lambda \eta
$$
 is a modification of the Korteweg-de Vries (KdV) equation for the study of  oceanic waves, which takes into account the effect of background rotation \cite{Ostrovsky}. It is well known that the rotation in the oceanographic problem ($\gamma \lambda >0$) eliminates the  solitary wave solutions of the Korteweg-de Vries equation through the terminal radiation damping \cite{Leonov,GHO}. The numerical simulations in Refs. \cite{Helfrich, GH} have shown that a localised wave packet emerges as a stable dominant solution of the Ostrovsky equation.  In an independent study \cite{YK}, it was established that stable envelope solitons play a central role in the dynamics of a modified Toda lattice with an additional linear term, which can be related to the two-directional generalisation of the Ostrovsky equation derived in Ref. \cite{Gerkema}. The weakly nonlinear description of the emerging wave packet for the Ostrovsky equation in terms of a higher-order nonlinear Schr\"odinger equation has been developed in Ref. \cite{GH}, linking the wavenumber of the carrier wave with the extremum of the group velocity \cite{GH,YK}.
 
 In this paper, we are concerned with the construction of a weakly nonlinear solution of the initial-value problem for a system of coupled regularised Boussinesq  (cRB) equations  \cite{KSZ}
  \begin{eqnarray}
&&u_{tt} - u_{xx} =  u_x u_{xx} + u_{ttxx} - \delta (u-w), \nonumber \\
&&w_{tt} - c^2 w_{xx} = \alpha  w_x w_{xx} + \beta  w_{ttxx} + \gamma   (u-w). \qquad
\label{uw}
\end{eqnarray}
The regularised version of the Boussinesq equation is preferable from the viewpoint of numerical simulations due to suppression of the short wave instability (see Refs. \cite{BBM, Christov}), although this version is not integrable by the Inverse Scattering Transform \cite{Z, AS}. From the viewpoint of our developed analytical approach, this is not essential, and we could have worked with any version of coupled Boussinesq equations. Also, within our approach, generalisation of the derivations to the case of three and more equations of this type is straightforward, and we do not discuss it in this paper, although the detailed study of various physical effects  is interesting. 
In the context of waves in solids, Boussinesq-type equations  have been derived, for example, for nonlinear waves in solid waveguides  \cite{Samsonov1,PS,Samsonov2} and for waves in microstructured solids \cite{Porubov, JE}  (further references can be found in Ref. \cite{Maugin} and Refs. \cite{C1,C2,C3}).

System (\ref{uw}) has been recently derived as an accurate asymptotic model for long nonlinear longitudinal waves in a layered waveguide with a soft bonding layer using a complex nonlinear layered lattice model  \cite{KSZ}. The system may be also obtained as  a continuum limit for a model of two one-dimensional Fermi-Pasta-Ulam (FPU)  chains \cite{FPU} with weak coupling between them.  System (\ref{uw}) is Lagrangian with the Lagrangian density
\begin{eqnarray*}
L &=& \frac12 \left [u_t^2 + \frac{\delta}{\gamma} w_t^2 - u_x^2 - \frac{\delta c^2}{\gamma} w_x^2 - \frac 13 \left (u_x^3 + \frac{\alpha \delta}{\gamma} w_x^3 \right )\right. 
+  \left. u_{tx}^2 + 
\frac{\beta \delta}{\gamma} w_{tx}^2 - \delta (u-w)^2  \right ].
\end{eqnarray*}
It has three conservation laws $A^i_t + B^i_x = 0, i=1,2,3$, with the densities
\begin{eqnarray*}
&&A^1 = u_t + \frac{\delta}{\gamma} w_t , \\
&&A^2 =  \frac12 \left [u_t^2 + \frac{\delta}{\gamma} w_t^2 + u_x^2 + \frac{\delta c^2}{\gamma} w_x^2 + \frac 13 \left (u_x^3 + \frac{\alpha \delta}{\gamma} w_x^3 \right )+ u_{tx}^2 +
\frac{\beta \delta}{\gamma} w_{tx}^2 + \delta (u-w)^2  \right ],\\
&&A^3 = u_t u_x + \frac{\delta}{\gamma} w_t w_x + u_{tx} u_{xx} + \frac{\beta \delta}{\gamma} w_{tx} w_{xx}
\end{eqnarray*}
 (for details see Ref. \cite{KSZ}).
%\begin{equation}
%A^i_t + B^i_x = 0, \quad i = 1,2,
%\label{cl}
%\end{equation}
%where the respective densities and flows are given by
%\begin{eqnarray*}
%A^1 &=& \frac12 \left [u_t^2 + \frac{\delta}{\gamma} w_t^2 + u_x^2 + \frac{\delta c^2}{\gamma} w_x^2 + \frac 13 \left (u_x^3 + \frac{\alpha \delta}%{\gamma} w_x^3 \right )\right. 
%+  \left. u_{tx}^2 + 
%\frac{\beta \delta}{\gamma} w_{tx}^2 + \delta (u-w)^2  \right ], \\
%B^1 &=& -u_t u_x - \frac{\delta c^2}{\gamma} w_t w_x - \frac 12 u_t u_x^2 - \frac{\alpha \delta}{2 \gamma} w_t w_x^2 - u_t u_{ttx} 
%- \frac{\beta \delta}{\gamma} w_t w_{ttx}; \\
%A^2 &=& u_t u_x + \frac{\delta}{\gamma} w_t w_x + u_{tx} u_{xx} + \frac{\beta \delta}{\gamma} w_{tx} w_{xx}, \\
%B^2 &=& -\frac12 \left [u_t^2 + \frac{\delta}{\gamma} w_t^2 + u_x^2 + \frac{\delta c^2}{\gamma} w_x^2  +
% \frac 23 \left (u_x^3 + \frac{\alpha \delta}{\gamma} w_x^3 \right ) \right .
% + \left . u_{tx}^2 + \frac{\beta \delta}{\gamma} w_{tx}^2 - \delta (u-w)^2 \right ] \\
% &-& u_x u_{ttx} - \frac{\beta \delta}{\gamma} w_x w_{ttx}.
%\end{eqnarray*}
%Indeed, one can verify that equations (\ref{cl}) hold by virtue of equations (\ref{uw}).

Differentiating (\ref{uw}) with respect to $x$, and denoting $u_x = f, w_x = g$, we obtain 
\begin{eqnarray}
&&f_{tt} - f_{xx} = \frac 12 (f^2)_{xx} + f_{ttxx} - \delta (f-g), \nonumber \\
&&g_{tt} - c^2 g_{xx} = \frac 12  \alpha  (g^2)_{xx} + \beta  g_{ttxx} + \gamma  (f-g) \qquad
\label{fg}
\end{eqnarray}
(uncoupled equations in this form are sometimes called  ``regularised long wave equations" and  ``improved bad Boussinesq equations").
In what follows we will consider solutions of this system instead of the solutions of the original system (\ref{uw}).  We refer to both systems (\ref{uw}) and (\ref{fg}) as coupled regularised Boussinesq (cRB) equations, since system (\ref{fg}) is obtained by differentiation of system (\ref{uw}). 

We are interested in constructing a weakly nonlinear solution of the initial-value problem (IVP) for  system (\ref{fg}).
We use asymptotic multiple-scales expansions of the type used in the study of oblique interaction of solitary waves in Refs. \cite{Miles1,Miles2} (see also references therein and in Ref. \cite{GG}). Recently, we developed a scheme based on this type of asymptotic expansions, which allowed us to solve a weakly nonlinear wave scattering problem \cite{Fission}, formulated in terms of a Boussinesq-type equation with piecewise-constant coefficients subject to two continuity conditions across the jump and some natural radiation conditions.  In this paper, we first use the procedure of averaging with respect to the fast time to obtain a hierarchy of asymptotically exact coupled and uncoupled Ostrovsky equations for the cases when the characteristic linear speeds of the two wave operators in Eqs.  (\ref{fg}) are close or essentially different ({\it strong} or {\it weak} interactions in the terminology of Refs. \cite{Miles1,Miles2,GG}). More precisely, to leading order we derive four uncoupled Ostrovsky equations when $c-1 = O(1)$, but two coupled systems of Ostrovsky equations when $c-1 = O(\epsilon)$. 
 Then, we show how to construct the weakly nonlinear solution of the IVP in terms of solutions of the derived Ostrovsky equations within the accuracy of the Eqs. (\ref{fg}). We also establish that corrections to the leading-order terms are nonsecular due to a special property of solutions of the Ostrovsky equation. In the absence of  coupling ($\delta = \gamma = 0$), these results yield a weakly nonlinear solution of the initial-value problem for the Boussinesq equation in terms of solutions of the initial-value problems for two Korteweg-de Vries  (KdV) equations. Finally, we perform numerical simulations of the original unapproximated system (\ref{fg}) to show the difference in the asymptotic behaviour of its solutions, when initial conditions are taken in the form of co-propagating solitary waves of the uncoupled equations. The results support our theory.
 
 The word `hierarchy' is used here to reflect on the growing complexity of the leading order asymptotic models.
% More precisely, to leading order we derive four uncoupled Ostrovsky equations when $c-1 = O(1)$, but two coupled systems of Ostrovsky equations when $c-1 = O(\epsilon)$. Note that generalisation to the case of $N$ coupled Boussinesq-type equations (describing, for example, long longitudinal waves in the $N$-layered elastic waveguide, or waves in $N$ coupled FPU chains), with $N$ different characteristic speeds, will lead to even greater number of different asymptotic models of growing complexity: from $2N$ uncoupled Ostrovsky equations to two systems of $N$ coupled Ostrovsky equations, and various intermediate cases.
In particular, generalisation to the case of $N$ coupled Boussinesq-type equations (describing, for example, long longitudinal waves in  $N$-layered elastic waveguide, or waves in $N$ coupled FPU chains), with $N$ characteristic speeds close to each other, will lead to a system of $N$ coupled Ostrovsky equations.

\section{Dispersion curve and solitary waves}

In the symmetric case, when $c=\alpha=\beta=1$, system (\ref{fg}) admits a reduction $g=f$, where $f$ satisfies the equation
\begin{equation}
f_{tt} - f_{xx} = \frac 12 (f^2)_{xx} + f_{ttxx}.
\label{f}
\end{equation}

Eq. (\ref{f}) has particular solutions in the form of  solitary waves:
%\begin{equation}
$$
f = A\  {\rm sech}^2 \frac{x - v t}{\Lambda},
\label{soliton}
$$
%\end{equation}
where $A = 3 (v^2 - 1), \ \Lambda = \frac{2 v}{\sqrt{v^2 - 1}}$. 
%shown in Fig.\ref{CSW} for $v = 1.2$.
%\begin{figure}[h]
%\begin{center}
%\includegraphics[width=5.5cm]{Fig5a} \
%\includegraphics[width=5cm]{Fig5b}
%\caption{\small Solitary wave (\ref{soliton}) for $v=1.2$.}
%\label{CSW}
%\end{center}
%\end{figure}
However,  these {\it pure} or {\it classical} solitary wave solutions, rapidly decaying to zero in their tail regions, are structurally unstable and are replaced with {\it radiating} solitary waves \cite{KSZ}, i.e. a solitary wave radiating a co-propagating one-sided oscillatory tail, using  the terminology of Refs. \cite{BGK, Shrira, Bona}. There have been extensive studies of generalised and radiating solitary waves, especially in the context of fluid mechanics (e.g., \cite{VB} - \cite{Grimshaw}),
%  Karpman, GJ, Boyd, Lombardi, Ch, GI, Grimshaw}).
but the models were different from Eqs. (\ref{fg}). In particular, long-wave ripples are radiated by solitons in Eqs.  (\ref{fg}), due to the type of coupling terms in the equations, and the resulting structure of the dispersion relation.

\begin{figure}[h]
\begin{center}
\includegraphics[width=5.3cm]{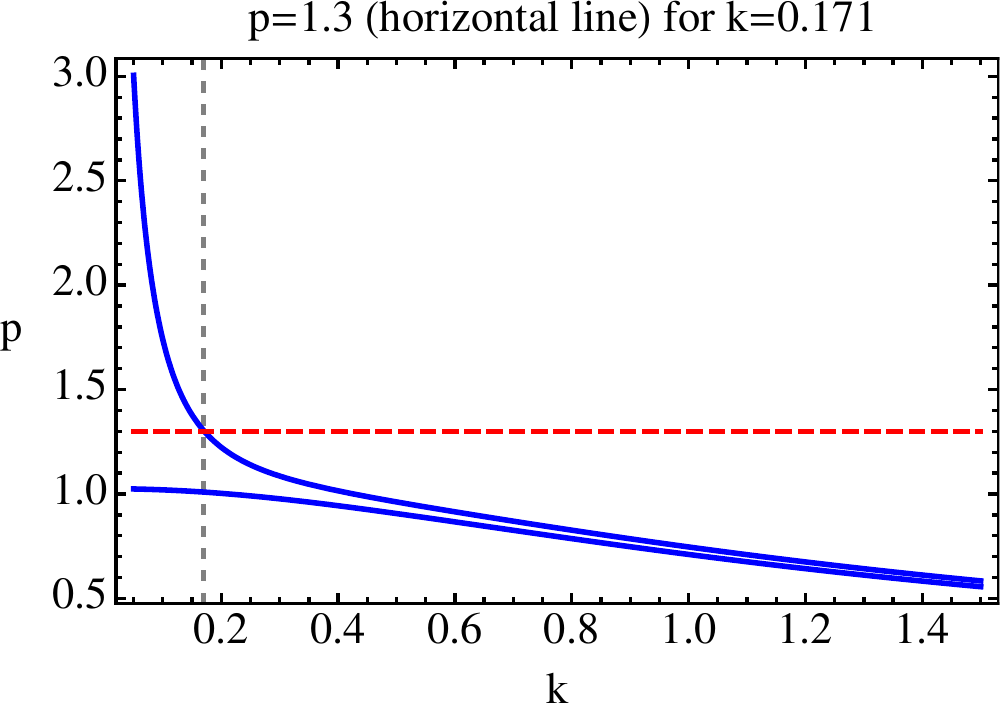} \quad
\includegraphics[width=5.3cm]{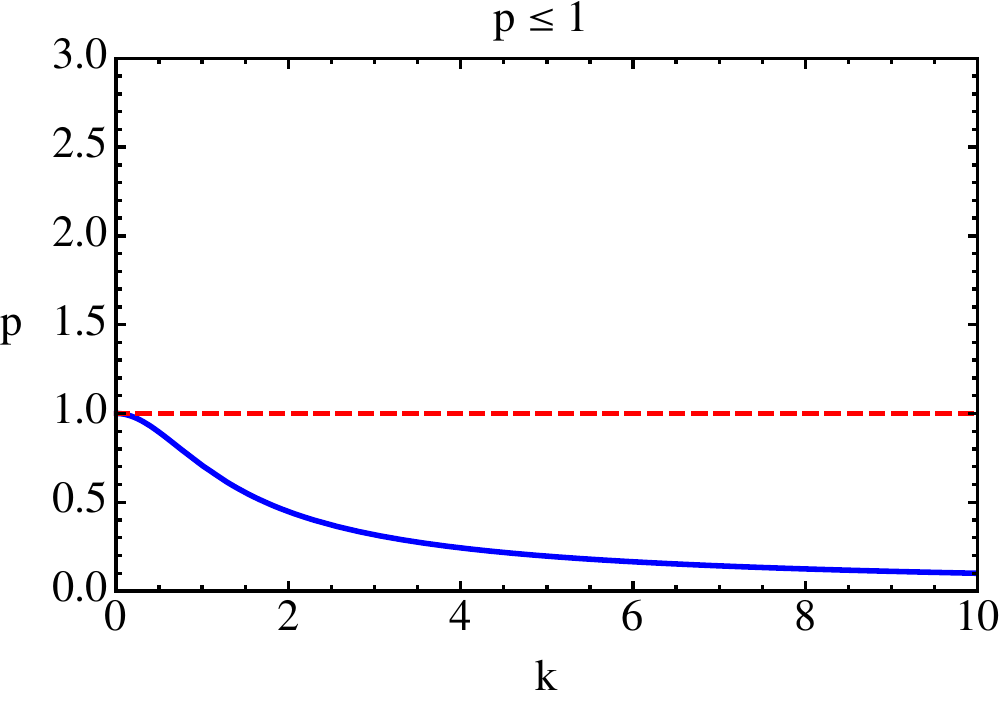}
\caption{\small (a) Two branches of the linear dispersion curve of Eqs. (\ref{fg}) for $c=1.05, \beta = 1,  \delta = \gamma =0.01$ and intersection with $p= 1.3$ (horizontal line) and (b) Linear dispersion curve of the reduction (\ref{f})  in the symmetric case $c=1, \beta = 1,  \delta = \gamma =0.01$.
%(left) and $c=1.05, \beta=1.05, \delta = \gamma =0.1$ (right) 
}
\label{disp_curves}
\end{center}
\end{figure}

We consider system (\ref{fg}), and assume that coefficients are perturbed compared to the symmetric case above, but remain positive.  The linear dispersion relation has the form
%\begin{equation}
$$
[k^2 (1-p^2) - k^4 p^2 + \delta] [k^2 (c^2-p^2) - \beta k^4 p^2 + \gamma] = \gamma \delta,
$$
%\label{disp}
%\end{equation}
where $k$ is the wavenumber and $p$ is the phase speed, 
and was analysed in Ref. \cite{KSZ}. A typical linear dispersion curve of Eqs. (\ref{fg}) is shown in Fig.\ref{disp_curves}. Significant difference with the linear dispersion curve of the reduction (\ref{f})  consists in the appearance of the second (upper) branch, going to infinity as $k \to 0$, and approaching zero, remaining above the lower branch, as $k \to \infty$. 
\medskip

%\vspace{-0.5cm}

The {\it classical} or {\it pure} solitary waves of the single Boussinesq equation (\ref{f})  arise as a bifurcation from wavenumber $k = 0$ of the linear wave spectrum, shown in Fig.\ref{disp_curves} (b), when there is no possible resonance with any linear wave for any value of $k$. The solitary wave speed $v$  is greater than  the linear long wave speed, i.e. $v > 1$, while the speed of a linear wave of any wavenumber is smaller, i.e. $p \le 1$.  This becomes generally impossible when the symmetry is broken. Instead, {\it radiating} solitary waves  arise  for the case when there is a possible resonance with the upper branch for some finite non-zero value of $k$. For example, a possible resonance is shown in Fig.\ref{disp_curves} (a) for $v=p=1.3$.  The  solitary wave solutions of Eq. (\ref{f}), viewed as particular solutions of the coupled equations in the symmetric case, constitute a one-parameter family of so-called {\it embedded} solitary waves (e.g., \cite{Ch,Yang}). Recently, radiating solitary waves have been experimentally observed in two- and three-layered elastic waveguides with soft bonding layers \cite{DSSK}.

From the studies of solitary waves in Refs. \cite{Miles1,Miles2,GG} and our recent studies of the dynamics of weakly nonlinear wave packets \cite{GGK} we know that the dynamics and the asymptotic models depend on the relative speeds of the waves. A question arises, to what extent does the difference between the characteristic linear speeds of the two wave operators (i.e., $c-1$) effect the dynamics of the nonlinear waves in Eqs. (\ref{fg})? In particular, if we take the initial conditions in the form of the solitary waves of the uncoupled Boussinesq equations, will the outcome be different for the cases $c-1 = O(\epsilon)$ and $c-1 = O(1)$, where $\epsilon$ is the natural small parameter of the Boussinesq model? The following analysis shows that this difference is crucial.

\section{Weakly nonlinear solution of the initial-value problem}

In this section we are concerned with unidirectional waves, which constitute the leading-order terms in our asymptotic multiple-scales expansions. We  use an averaging procedure which allows us to derive asymptotic reductions to simpler asymptotically exact models in the form of coupled and uncoupled Ostrovsky equations, and construct a weakly nonlinear solution of the initial-value problem in terms of solutions of the derived Ostrovsky equations.

For these purposes we need to rewrite system (\ref{fg}) in the original unscaled form (see Ref. \cite{KSZ}), substituting
$$
f =  \epsilon  \tilde f , \quad g = \epsilon  \tilde g, \quad \tilde t = \sqrt{\epsilon}\  t, \quad \tilde x = \sqrt{\epsilon}\  x, \quad \delta = \epsilon^2 \tilde \delta,\quad  \gamma = \epsilon^2 \tilde \gamma
$$
into the system (\ref{fg}), to obtain (omitting the tildes)
\begin{eqnarray}
&& f_{tt}- f_{xx} = \epsilon \left [\frac 12 (f^2)_{xx} + f_{ttxx} - \delta (f-g)\right ], \nonumber \\
&& g_{tt}-c^2 g_{xx} = \epsilon \left [\frac 12 \alpha (g^2)_{xx}  +  \beta g_{ttxx} + \gamma (f-g)\right ].
\label{fg1}
\end{eqnarray}
As any Boussinesq-type  system, system (\ref{fg1}) appears as an approximation containing $O(1)$ and $O(\epsilon)$ terms (see Ref. \cite{KSZ}).

We consider the Cauchy problem for Eq. (\ref{fg1}) on the infinite line, imposing the following initial conditions:
\begin{eqnarray}
&&f|_{t=0} = F(x), \quad g|_{t=0} = G(x), \label{IC1}\\
&&f_t|_{t=0} = V(x), \quad g_t|_{t=0} = W(x). \label{IC2}
\end{eqnarray}
Some local existence results applicable to this problem were recently obtained in Ref. \cite{Duruk} (Theorem 2.4 and Remark 2.5, according to  Ref. \cite{Duruk1}). In this paper, we are concerned with the {\it explicit}  construction of the weakly nonlinear solution of the Cauchy problem in terms of the asymptotically exact (KdV-like) models for unidirectional waves.

We assume that the initial conditions are sufficiently rapidly decaying at both infinities, so that to leading order the initial  ($t = O(1)$) evolution  of the Cauchy data is described by the classical D'Alembert's solution
$$
f_0 (t,x) = f_0^- (x-t) + f_0^+ (x+t), \quad g_0 (t,x) = g_0^- (x-ct) + g_0^+ (x+ct),
$$
where
\begin{eqnarray}
&& f_0^{\pm} (x \pm t) = \frac 12 \left ( F(x \pm t) \pm \int_{-\infty}^{x \pm t} V(x) dx \right ), \label{f0} \\
&& g_0^{\pm} (x \pm c t) = \frac 12 \left ( G(x \pm c t) \pm \frac 1c \int_{-\infty}^{x \pm c t} W(x) dx \right ). \label{g0}
\end{eqnarray}
In general, $f^\pm_0$ and $g^\pm_0$ are some step-like functions. In what follows we restrict our considerations to the case when these functions are sufficiently rapidly decaying at infinity (i.e. $\int_{-\infty}^{\infty} V(x) dx = 0$ and $\int_{-\infty}^{\infty} W(x) dx = 0$).

To describe the subsequent ($t  = O(\epsilon^{-1})$)  evolution of the given initial data we introduce the slow time $T = \epsilon t$ and look for the weakly nonlinear solution of the Cauchy problem (\ref{fg1}) -  (\ref{IC2}) in the form of asymptotic multiple-scales expansions. The form of these expansions depends on  the difference between the characteristic speeds of the linear wave operators in system (\ref{fg1}), and below we consider the two main cases, when $c-1 = O(\epsilon)$ and $c-1 = O(1)$.

\subsection{Case I: $c-1 = O(\epsilon)$}

In this case, we rewrite system (\ref{fg1}) as
\begin{eqnarray}
&& f_{tt} - f_{xx} = \epsilon \left [\frac 12 (f^2)_{xx} + f_{ttxx} - \delta (f-g) \right ], \nonumber \\
&& g_{tt} - g_{xx} = \epsilon \left [\frac 12 \alpha (g^2)_{xx}  +  \beta g_{ttxx} + \gamma (f-g) + \frac{c^2-1}{\epsilon} g_{xx}\right ],
\label{fg2}
\end{eqnarray}
where $\frac{c^2-1}{\epsilon} \sim O(1)$ since $c-1 = O(\epsilon)$,
and look for the solution in the form
\begin{eqnarray}
&& f = f^-(\xi, T) + f^+(\eta, T) + \epsilon f^1(\xi, \eta, T) + O(\epsilon^2), \nonumber \\
&& g = g^-(\xi, T) + g^+(\eta, T) + \epsilon g^1(\xi, \eta, T) + O(\epsilon^2). \label{exp1}
\end{eqnarray}
Here, $\xi = x-t, \eta = x+t, T = \epsilon t$, and we consider each wave in its own reference frame.  We will view the leading order approximation of the linear solution (\ref{f0}) - (\ref{g0}) (i.e., for $c-1 = O(\epsilon )$, $c$ is replaced with 1) as initial conditions for the functions $f^-, f^+, g^-, g^+$ with respect to the slow time $T$, i.e.
\begin{equation}
f^{\pm}|_{T=0} = f_0^{\pm}, \quad g^{\pm}|_{T=0} = g_0^{\pm} 
\label{ICO1}
\end{equation}
(this is later derived at leading order, when we substitute our asymptotic expansions into the initial conditions (\ref{IC1}) and (\ref{IC2})). 

Substituting the expansions (\ref{exp1}) into Eqs. (\ref{fg2}), we find that the equations are satisfied at leading order, while at $O(\epsilon)$ we obtain
\begin{eqnarray}
-4 f^1_{\xi \eta} &=& (2 f^-_T + f^- f^-_{\xi} + f^-_{\xi \xi \xi})_\xi 
 +  (-2 f^+_T  + f^+ f^+_{\eta} + f^+_{\eta \eta \eta} )_\eta  \nonumber \\
 &+& 2 f^-_\xi f^+_\eta  + f^+ f^-_{\xi \xi} + f^- f^+_{\eta \eta} - \delta (f^- + f^+ - g^- - g^+), \label{f1} \\
 -4 g^1_{\xi \eta} &=& (2 g^-_T + \alpha g^- g^-_{\xi} + \beta g^-_{\xi \xi \xi})_\xi 
+  (-2 g^+_T + \alpha g^+ g^+ _{\eta} + \beta g^+_{\eta \eta \eta})_\eta \nonumber  \\
&+&  \alpha (2 g^-_\xi g^+_\eta   + g^+ g^-_{\xi \xi} + g^- g^+_{\eta \eta}) + \gamma (f^- + f^+ - g^- - g^+) \nonumber \\
&+&  \frac{c^2-1}{\epsilon} (g^-_{\xi \xi} + g^+_{\eta \eta}). \label{g1}
\end{eqnarray}

We next average equations (\ref{f1}) and (\ref{g1}) with respect to the fast time $t$, considering
$$\lim_{\tau \to \infty} \frac{1}{\tau} \int_0^{\tau} \dots dt$$
 at constant $\xi$ or $\eta$, i.e. in the reference frame moving with the linear speed of the right- or left-propagating waves, respectively. Indeed, requiring that $f^1$, $g^1$ and their derivatives remain bounded (which is necessary to avoid the appearance of secular terms in expansions (\ref{exp1})), we see that, for example at constant $\xi$,
$$
\lim_{\tau \to \infty} \frac{1}{\tau} \int_0^\tau f^1_{\xi \eta} dt = \lim_{\tau \to \infty} \frac{1}{2\tau} \int_{\xi}^{\xi + 2\tau} f^1_{\xi \eta} d \eta =  \lim_{\tau \to \infty} \frac{1}{2\tau} \left [f^1_\xi \right ]_\xi^{\xi+2\tau} =0,
$$
and similarly for $g^1$, as well as for $f^1$ and $g^1$ at constant $\eta$, showing that the averaging results in zeros in the left-hand sides of Eqs. (\ref{f1}) and (\ref{g1}).  Similarly, assuming that functions $f^-, f^+, g^-, g^+$ and their derivatives remain bounded and sufficiently rapidly decaying at infinity for any fixed $T$ (the assumptions  are consistent with the relevant numerical experiments), and averaging the entire Eq. (\ref{f1}) with respect to $t$ at constant $\xi$, we obtain
\begin{eqnarray}
0 & = &\left (2 f^-_T +  f^- f^-_{\xi} + f^-_{\xi \xi \xi}\right )_\xi  - \delta (f^- - g^-) \nonumber \\
& + &  \lim_{\tau \to \infty} \frac{1}{\tau} \int_0^{\tau} [ (-2 f^+_T + f^+ f^+_{\eta} + f^+_{\eta \eta \eta})_\eta 
 + 2 f^-_\xi f^+_\eta  + f^+ f^-_{\xi \xi} + f^- f^+_{\eta \eta} - \delta (f^+ - g^+)] dt \nonumber \\
& = & \left (2 f^-_T +  f^- f^-_{\xi} + f^-_{\xi \xi \xi}\right )_\xi  - \delta (f^- - g^-) \nonumber \\
& + &  \lim_{\tau \to \infty} \frac{1}{2 \tau} \int_{\xi}^{\xi + 2\tau} [ (-2 f^+_T + f^+ f^+_{\eta} + f^+_{\eta \eta \eta})_\eta 
 + 2 f^-_\xi f^+_\eta  + f^+ f^-_{\xi \xi} + f^- f^+_{\eta \eta} - \delta (f^+ - g^+)] d \eta \nonumber \\
&  = &  \left (2 f^-_T +  f^- f^-_{\xi} + f^-_{\xi \xi \xi}\right )_\xi  - \delta (f^- - g^-),
 \label{O1a}
\end{eqnarray}
while averaging Eq. (\ref{g1}) at constant $\xi$ results in 
\begin{equation}
0 = \left (2 g^-_T + \alpha g^- g^-_{\xi} + \beta g^-_{\xi \xi \xi}\right )_\xi  + \gamma (f^- - g^-) + \frac{c^2-1}{\epsilon} g^-_{\xi \xi}.
\label{O1b}
\end{equation}
Similarly, averaging Eq. (\ref{f1}) at constant $\eta$ under the same assumptions yields
\begin{equation}
0 = \left (-2 f^+_T + f^+ f^+_{\eta} + f^+_{\eta \eta \eta} \right )_\eta - \delta (f^+ - g^+),
\label{O2a}
\end{equation}
while averaging Eq. (\ref{g1}) at constant $\eta$ results in 
\begin{equation}
0 = \left (-2 g^+_T + \alpha g^+ g^+_{\eta} + \beta g^+_{\eta \eta \eta}\right )_\eta + \gamma (f^+ - g^+) + \frac{c^2-1}{\epsilon} g^+_{\eta \eta}.
\label{O2b}
\end{equation}
Thus, to leading order we obtain two systems of coupled Ostrovsky equations. Next, substituting the Eqs. (\ref{O1a}) - (\ref{O2b})  back into Eqs. (\ref{f1}) and (\ref{g1}), we obtain equations for the higher-order corrections
$$
f^1_{\xi \eta} = - \frac 14 \left (2 f^-_\xi f^+_\eta + f^+ f^-_{\xi \xi} + f^- f^+_{\eta \eta} \right ), \quad g^1_{\xi \eta} = - \frac {\alpha}4 \left ( 2 g^-_\xi g^+_\eta +  g^+ g^-_{\xi \xi} + g^-  g^+_{\eta \eta} \right ),
$$
which imply
\begin{eqnarray*}
&&f^1 = - \frac 14 \left (2 f^- f^+  + f^-_{\xi} \int f^+ d\eta + f^+_{\eta} \int f^- d\xi  \right )+ \phi_1(\xi, T) + \psi_1(\eta, T), \\
&& g^1 = -\frac {\alpha}4 \left (2 g^- g^+   +  g^-_{\xi} \int g^+ d\eta + g^+_{\eta} \int g^- d\xi \right ) + \phi_2(\xi, T) + \psi_2(\eta, T).
\end{eqnarray*}
The presence of four arbitrary functions allows us to satisfy not only the equations, but also the initial conditions (\ref{IC1}) and (\ref{IC2}) up to $O(\epsilon^2)$, constructing therefore an accurate asymptotic solution of the initial-value problem (within the accuracy of the problem formulation). 

Indeed, substituting our expansions (\ref{exp1}) into the initial conditions (\ref{IC1}) and (\ref{IC2}) to leading order we recover the formulae  (\ref{ICO1}) for the initial conditions, while at $O(\epsilon)$ we obtain D'Alembert's-like formulae for the functions $\phi_i (\xi, T)$ and $\psi_i (\eta, T), i=1,2$:
\begin{eqnarray}
\phi_i(\xi, T) = \frac 12 \left [R_{i1} (\xi, T) + \int_{-\infty}^\xi R_{i2} (x, T) dx\right ], \nonumber \\
\psi_i(\eta, T) = \frac 12 \left [R_{i1} (\eta, T) - \int_{-\infty}^\eta R_{i2} (x, T) dx\right ], 
\label{phipsi}
\end{eqnarray}
where
\begin{eqnarray*}
&&R_{11} (x, T) = \frac 14 \left [2 f^{-} f^{+} + f^-_{\xi} \int f^+ d\eta + f^+_{\eta} \int f^- d\xi  \right ]_{t=0}, \\
&&R_{12} (x, T) = \left [f^{-}_T + f^{+}_T + \frac 14 \left (f^+ f^{-}_{\xi} - f^{-} f^{+}_{\eta} +  f^-_{\xi \xi} \int f^+ d\eta - f^+_{\eta \eta} \int f^- d\xi  \right  )\right ]_{t=0}, \\
&&R_{21} (x, T) = \frac {\alpha}4 \left [2 g^{-} g^{+}  + g^-_{\xi} \int g^+ d\eta + g^+_{\eta} \int g^- d\xi\right ]_{t=0}, \\
&&R_{22} (x, T) = \left [g^{-}_T + g^{+}_T + \frac {\alpha}4 \left (g^+ g^{-}_{\xi}  - g^{-} g^{+}_{\eta} +  g^-_{\xi \xi} \int g^+ d\eta - g^+_{\eta \eta} \int g^- d\xi  \right ) \right ]_{t=0}.
\end{eqnarray*}
Thus, within the accuracy of the problem formulation (i.e. $O(\epsilon^2)$), the dependence of functions $\phi$ and $\psi$ on the characteristic variables is determined, while their dependence on the slow time $T$ is inherited from the dependence of the leading order waves, or it may be neglected, at least for sufficiently small values of time.

The leading order systems of coupled Ostrovsky equations for unidirectional waves can be rewritten in a symmetric form in the reference frames moving with the average linear speed $\bar c = \frac{c+1}{2}$, i.e. formally changing $\xi$ and $\eta$ in (\ref{O1a}) - (\ref{O2b}) to $\bar \xi = \xi - \Delta {\bar c} T$ and $\bar \eta = \eta + \Delta {\bar c} T$, $\Delta = \frac{c-1}{2 \epsilon}$, which yields
\begin{eqnarray*}
&&\left [ 2 (f^{-}_T - \Delta {\bar c} f^{-}_{\bar \xi}) + f^{-} f^{-}_{\bar \xi} + f^{-}_{{\bar \xi}{\bar \xi}{\bar \xi}} \right ]_{\bar \xi}=\delta (f^{-} - g^{-}), \\
&&\left [ 2 (g^{-}_T + \Delta {\bar c} g^{-}_{\bar \xi}) + \alpha g^{-} g^{-}_{\bar \xi} + \beta g^{-}_{{\bar \xi}{\bar \xi}{\bar \xi}} \right ]_{\bar \xi}= - \gamma (f^{-} - g^{-}), 
\end{eqnarray*}
and
\begin{eqnarray*}
&&\left [ 2 (f^{+}_T + \Delta {\bar c} f^{+}_{\bar \eta}) - f^{+} f^{+}_{\bar \eta} - f^{+}_{{\bar \eta}{\bar \eta}{\bar \eta}} \right ]_{\bar \eta}=-\delta (f^{+} - g^{+}), \\
&&\left [ 2 (g^{+}_T - \Delta {\bar c} g^{+}_{\bar \eta}) - \alpha g^{+} g^{+}_{\bar \eta} - \beta g^{+}_{{\bar \eta}{\bar \eta}{\bar \eta}} \right ]_{\bar \eta}= \gamma (f^{+} - g^{+}). 
\end{eqnarray*}
Systems of coupled KdV equations have appeared in the literature before (see Refs. \cite{GG,Grimshaw} and references therein). To the best of our knowledge this is the first appearance of the coupled Ostrovsky equations.

\subsection{Case II: $c-1 = O(1)$}

In this case, we look for the solution in a different form:
\begin{eqnarray}
&& f = f^-(\xi_1, T) + f^+(\eta_1, T) + \epsilon f^1(\xi_1, \eta_1, T) + O(\epsilon^2), \nonumber \\
&& g = g^-(\xi_2, T) + g^+(\eta_2, T) + \epsilon g^1(\xi_2, \eta_2, T) + O(\epsilon^2), \label{exp2}
\end{eqnarray}
where $\xi_1 = x-t, \eta_1 = x+t,$ and $ \xi_2 = x-ct, \eta_2 = x+ct$ are the two pairs of characteristic variables for the two linear wave operators in Eqs. (\ref{fg1}), and again, we consider each wave in its own reference frame.

Substituting our expansions (\ref{exp2}) into Eqs. (\ref{fg1}) we obtain
\begin{eqnarray}
-4 f^1_{\xi_1 \eta_1} &=& (2 f^-_T + f^- f^-_{\xi_1} + f^-_{\xi_1 \xi_1 \xi_1})_{\xi_1} 
 +  (-2 f^+_T + f^+ f^+_{\eta_1} + f^+_{\eta_1 \eta_1 \eta_1})_{\eta_1}  \nonumber \\
 &+& 2 f^-_{\xi_1} f^+_{\eta_1}  + f^+ f^-_{\xi_1 \xi_1} + f^- f^+_{\eta_1 \eta_1}  - \delta (f^- + f^+ - g^- - g^+), \label{f1a} \\
 -4 c^2 g^1_{\xi_2 \eta_2} &=& (2c g^-_T + \alpha g^- g^-_{\xi_2} + \beta c^2 g^-_{\xi_2 \xi_2 \xi_2})_{\xi_2} 
+  (-2c g^+_T + \alpha g^+ g^+_{\eta_2} + \beta c^2 g^+_{\eta_2 \eta_2 \eta_2})_{\eta_2} \nonumber  \\
&+& \alpha (2 g^-_{\xi_2} g^+_{\eta_2}  + g^+ g^-_{\xi_2 \xi_2} +  g^- g^+_{\eta_2 \eta_2})  + \gamma (f^- + f^+ - g^- - g^+). \label{g1a}
\end{eqnarray}

Under the same assumptions as before, we can average Eqs. (\ref{f1a}) and (\ref{g1a}) with respect to $t$, considering 
$$\lim_{\tau \to \infty} \frac{1}{\tau} \int_0^{\tau} \dots dt$$
  at constant $\xi_1$ or $\eta_1$, and $\xi_2$ or $\eta_2$, respectively, obtaining in this case four uncoupled Ostrovsky equations:
\begin{eqnarray}
 (2 f^-_T + f^- f^-_{\xi_1} + f^-_{\xi_1 \xi_1 \xi_1})_{\xi_1} = \delta f^-, \label{eqn: one(f-) } \label{O1}\\
(2 f^+_T - f^+ f^+_{\eta_1} - f^+_{\eta_1 \eta_1 \eta_1})_{\eta_1}  =- \delta f^+,  \label{O2}\\
(2c g^-_T + \alpha g^- g^-_{\xi_2} + \beta c^2 g^-_{\xi_2 \xi_2 \xi_2})_{\xi_2} = \gamma g^-,  \label{eqn: one(g-) } \label{O3}\\
(2c g^+_T - \alpha g^+ g^+_{\eta_2} - \beta c^2 g^+_{\eta_2 \eta_2 \eta_2})_{\eta_2} =- \gamma g^+, \label{O4}
\end{eqnarray}
and equations for the higher-order corrections
\begin{eqnarray}
 f^1_{\xi_1 \eta_1} = - \frac 14 \left (2 f^-_{\xi_1}  f^+_{\eta_1}  +  f^+ f^-_{\xi_1 \xi_1} +  f^- f^+_{\eta_1 \eta_1}   \right ) - \frac{\delta}{4} (g^- + g^+), \label{f1b}\\
 g^1_{\xi_2 \eta_2} = - \frac {\alpha}{4 c^2} \left (2 g^-_{\xi_2} g^+_{\eta_2} + g^+ g^-_{\xi_2 \xi_2} +  g^- g^+_{\eta_2 \eta_2}  \right ) - \frac{\gamma}{4 c^2}  (f^- + f^+), \label{g1b}
\end{eqnarray}
where in the right-hand sides we have solutions of the leading order Ostrovsky equations
$$
g^-(\xi_2, T) = g^- \left ( \frac{(1+c) \xi_1 + (1-c) \eta_1}{2}, T \right ), \quad g^+(\eta_2, T) = g^+ \left ( \frac{(1-c) \xi_1 + (1+c) \eta_1}{2}, T \right ),
$$
and 
$$f^-(\xi_1, T) = f^{-} \left ( \frac{(c+1) \xi_2 + (c-1) \eta_2}{2c}, T \right ), \quad f^+(\eta_1, T) = f^+ \left ( \frac{(c-1) \xi_2 + (c+1) \eta_2}{2c}, T \right ).
$$

Remarkably, the particular solutions of Eqs. (\ref{f1b}) and (\ref{g1b}) are bounded functions, because of the special property of the smooth  solutions of the Ostrovsky equation, namely
\begin{equation}
\int_{-\infty}^{\infty} f^- d\xi_1 = 0, \quad \int_{-\infty}^{\infty} f^+ d\eta_1 = 0, \quad \int_{-\infty}^{\infty} g^- d\xi_2 = 0, \quad \int_{-\infty}^{\infty} g^+ d\eta_2 = 0.
\label{mass}
\end{equation}
Indeed, the solution of  Eqs. (\ref{f1b}) and (\ref{g1b}) can be found in the form
\begin{eqnarray*}
f^1 =  - \frac 14 \left (2 f^- f^+ + f^-_{\xi_1} \int f^+ d\eta_1 + f^+_{\eta_1} \int f^- d\xi_1 \right )+ f_{p1}(\xi_2, T) + f_{p2}(\eta_2, T) + \phi_1(\xi_1, T) + \psi_1(\eta_1, T), \\
g^1 = - \frac {\alpha}{4 c^2} \left (2 g^- g^+ + g^-_{\xi_2} \int g^+ d\eta_2 + g^+_{\eta_2} \int g^- d\xi_2 \right ) + g_{p1}(\xi_1, T) + g_{p2}(\eta_1, T) + \phi_2 (\xi_2, T) + \psi_2 (\eta_2, T), 
\end{eqnarray*}
where
\begin{eqnarray*}
&&f_{p1} = \frac{\delta}{c^2-1} \int_{-\infty}^{\xi_2} \int_{-\infty}^{v} g^-(u, T) du dv, \quad 
f_{p2} = \frac{\delta}{c^2-1} \int_{-\infty}^{\eta_2} \int_{-\infty}^{v} g^+(u, T) du dv, \\
&&g_{p1} = -\frac{\gamma}{c^2-1} \int_{-\infty}^{\xi_1} \int_{-\infty}^{v} f^-(u, T) du dv, \quad 
g_{p2} = -\frac{\gamma}{c^2-1} \int_{-\infty}^{\eta_1} \int_{-\infty}^{v} f^+(u, T) du dv.
\end{eqnarray*}
Let us consider $f_{p1}$, for example. Here, $\int_{-\infty}^{\infty} g^{-}(u, T) du = 0$, because of the mentioned property of the Ostrovsky equation, immediately obtained by integrating Eq. (\ref{O3}). Moreover, using Eq. (\ref{O3}) and recalling that solutions are decaying at infinity, we obtain
\begin{eqnarray*}
\int_{-\infty}^{\infty} \int_{-\infty}^{v} g^-(u, T) du dv &=& \frac {1}{\gamma} \int_{-\infty}^{\infty} (2c g^-_T + \alpha g^- g^-_v + \beta c^2 g^-_{vvv}) dv \\
&=& \frac{2c}{\gamma}\  \frac{d}{dT} \int_{-\infty}^{\infty} g^-(v, T) dv = 0.
\end{eqnarray*}
Therefore, the particular solution $f_{p1}$ is a bounded function, and $\lim_{\xi_2 \to \pm \infty} f_{p1} = 0$. Similarly, other particular solutions are also bounded functions and there are no secular terms.

The presence of four arbitrary functions  allows us to satisfy the initial conditions with the desired accuracy, as in the previous case. Substituting our expansions (\ref{exp2}) into the initial conditions (\ref{IC1}) and (\ref{IC2}) to leading order we again recover  formulae (\ref{ICO1}), while at $O(\epsilon$) we obtain D'Alembert's-like formulae for $\phi_i (\xi_i, T)$ and $\psi_i (\eta_i, T), i=1,2$:
\begin{eqnarray}
\phi_i(\xi_i, T) = \frac 12 \left [R_{i1} (\xi_i, T) + \frac {1}{c_i} \int_{-\infty}^{\xi_i} R_{i2} (x, T) dx\right ], \nonumber \\
\psi_i(\eta_i, T) = \frac 12 \left [R_{i1} (\eta_i, T) - \frac {1}{c_i} \int_{-\infty}^{\eta_i} R_{i2} (x, T) dx\right ], 
\label{phipsinew}
\end{eqnarray}
where $c_1 = 1, c_2 = c$ and 
\begin{eqnarray*}
R_{11} (x, T) &=& \frac 14 \left [2 f^{-} f^{+}  + f^-_{\xi_1} \int f^+ d\eta_1 + f^+_{\eta_1} \int f^- d\xi_1 \right ]_{t=0} \\
& - & \frac{\delta}{c^2-1} \int_{-\infty}^x \int_{-\infty}^v \left [ g^-(u,T) + g^+(u,T) \right ] du dv, \\
R_{12} (x, T) &=& \left [f^{-}_T + f^{+}_T + \frac 14 \left  (f^+ f^{-}_{\xi_1}  - f^{-} f^{+}_{\eta_1} + f^-_{\xi_1 \xi_1} \int f^+ d \eta_1 - f^+_{\eta_1 \eta_1}  \int f^- d \xi_1 \right ) \right ]_{t=0} \\
& -  & \frac{\delta c}{c^2-1} \int_{-\infty}^x \left [ g^-(u,T) - g^+(u,T) \right ] du,\\
R_{21} (x, T) &=& \frac {\alpha}{4 c^2} \left [2 g^{-} g^{+} +  g^-_{\xi_2} \int g^+ d\eta_2 + g^+_{\eta_2} \int g^- d\xi_2  \right ]_{t=0} \\
& + & \frac{\gamma}{c^2-1} \int_{-\infty}^x \int_{-\infty}^v \left [ f^-(u,T) + f^+(u,T) \right ] du dv, \\
R_{22} (x, T) &=& \left [g^{-}_T + g^{+}_T + \frac {\alpha}{4 c} \left (g^+ g^{-}_{\xi_2} - g^{-} g^{+}_{\eta_2} + g^-_{\xi_2 \xi_2} \int g^+ d  \eta_2 - g^+_{\eta_2 \eta_2}  \int g^- d \xi_2 \right ) \right ]_{t=0} \\
& + & \frac{\gamma }{c^2-1} \int_{-\infty}^x \left [ f^-(u,T) - f^+(u,T) \right ] du.
\end{eqnarray*}

Thus, in both cases $c-1 = O(\epsilon)$ and $c-1 = O(1)$, the asymptotic multiple-scales expansions and the averaging procedure described above have allowed us to construct  nonsecular weakly nonlinear solutions of the given initial-value problem for the values of time up to $O(\epsilon^{-1})$, within the accuracy of the problem formulation. To construct a more accurate solution, and for greater values of time,  one would need to know higher-order terms in the original cRB Eqs. (\ref{fg1}). However, the derived hierarchy of Ostrovsky equations will still describe leading order terms in these expansions, making the study of the long-time evolution of its solutions interesting. 

To finish this section, we would like to make an important comment that although smooth solutions of the Ostrovsky equation must satisfy the {\it zero mass} constraints (\ref{mass}) (and similar conditions $\int_{-\infty}^{\infty} (f^- - g^-) \ d \bar \xi = 0$, etc. in the case of coupled Ostrovsky equations of section 3.1), this does not impose any forbidding restrictions on the choice of the initial conditions (\ref{IC1}) and (\ref{IC2}) for the cRB system (\ref{fg1}). Indeed, initial conditions for the Ostrovsky equation can always be modified by adding a long but very small amplitude (i.e. $O(\epsilon^2$) or smaller) `pedestal' (e.g., $O(\epsilon^n)$  constant over the finite $O(\epsilon^{-n})$  interval, $n \ge 2$), so that the composite initial condition has zero mass, but this does not lower the accuracy of the asymptotic solution; and a transition to the zero mass solution is very fast (see the discussion in  Ref. \cite{GM}). Numerical simulations for the Ostrovsky equation show that if a smooth initial condition has a nonzero mass, numerically the solution adjusts immediately, since this initial condition can be viewed as an approximation to the composite zero mass solution (with any given accuracy). Note that similar issues appear in connection with several other equations, for example, the Kadomtsev-Petviashvili equation (see Ref. \cite{AW}) and, more recently, the short-pulse equation (see Ref. \cite{Horikis}), where the notion of the {\it initial time layer} has been introduced to describe such transitions \cite{AW} (see also Ref. \cite{GM} and references therein).

%%%%%%%%%%%%%%%%%%%%%%%% NUMEIRCAL SIMULATIONS %%%%%%%%%%%%%%%%%%%%%%

\section{Numerical simulations}

In this section we discuss numerical simulations of solutions of system (\ref{fg}). We implement a finite difference scheme which is an extension of the scheme developed in Ref. \cite{soerensen} for a single regularised Boussinesq equation. Our emphasis in this section is to compare the numerical solutions for the two cases discussed in the previous section, i.e. when the difference in the characteristic speeds of the system is of $O(1)$ or $O(\epsilon)$.

We let $x\in [-L,L]$, for finite $L$, and discretise the $(x,t)$ domain into a grid with spacings $\Delta x=h$ and $\Delta t=k$. The solutions $f(x,t)$ and $g(x,t)$ of Eqs. (\ref{fg}) are approximated by the solution $f(ih,jk)$ and $g(ih,jk)$ (for $i=0,1,...,N$ and $j=0,1,...$) of the difference scheme, denoted $f_{i,j}$ and $g_{i,j}$. 

Substituting central difference approximations into  system  (\ref{fg}) we derive the following difference schemes for the 
%two equations in 
system (\ref{fg}):
\begin{eqnarray}
 - f_{i-1,j+1} + (2 +h^2)f_{i,j+1}-f_{i+1,j+1} &=& (k^2-2)[f_{i-1,j}-2f_{i,j}+f_{i+1,j}] + 2h^2f_{i,j} 
 \nonumber \\
&& + \frac{k^2}{2}[(f_{i-1,j})^2-2(f_{i,j})^2+(f_{i+1,j})^2]
\nonumber
\\
&& + f_{i-1,j-1} - (2 +h^2)f_{i,j-1}+ f_{i+1,j-1}\nonumber \\
&& -h^2k^2\delta(f_{i,j} - g_{i,j}),
\label{eqn: 2layer scheme a} \\ \nonumber \\
 -\beta g_{i-1,j+1} + (2\beta +h^2)g_{i,j+1}-\beta g_{i+1,j+1} &=& (k^2c^2-2\beta)[g_{i-1,j}-2g_{i,j}+g_{i+1,j}] + 2h^2g_{i,j} 
 \nonumber \\
&& + \frac{\alpha k^2}{2}[(g_{i-1,j})^2-2(g_{i,j})^2+(g_{i+1,j})^2]
\nonumber
\\
&& +\beta g_{i-1,j-1} - (2\beta +h^2)g_{i,j-1}+\beta g_{i+1,j-1}\nonumber \\
&& + h^2k^2\gamma(f_{i,j} - g_{i,j}) .
\label{eqn: 2layer scheme b}
\end{eqnarray}

The boundary conditions are imposed far enough from the propagating waves, thus we set
\begin{eqnarray}
f_{0,j}=f_{N,j}=g_{0,j}=g_{N,j}=0, \quad \forall j.
\label{eqn: 2layer BC}
\end{eqnarray}
The initial conditions are chosen in the form of the co-propagating pure solitary wave solutions of the uncoupled equations ($\delta=\gamma=0$):
\begin{eqnarray}
f_{i,0} &=& A_1\  \text{sech} ^2\left(\frac{x}{\Lambda_1}\right), \qquad 
f_{i,1} = A_1\  \text{sech} ^2\left(\frac{x-v_1k}{\Lambda_1}\right), \nonumber \\
g_{i,0} &=& A_2\  \text{sech} ^2\left(\frac{x}{\Lambda_2}\right), \qquad
g_{i,1} =A_2\  \text{sech} ^2\left(\frac{x-v_2k}{\Lambda_2}\right), \quad \forall i,
\label{eqn: 2layer IC}
\end{eqnarray}
where $A_1=3(v_1^2-1)$, $A_2=\frac{3}{\alpha}(v_2^2-c^2)$, $\Lambda_1=2v_1(v_1^2-1)^{-\frac{1}{2}}$ and $\Lambda_2=2v_2\sqrt{\beta}(v_2^2-c^2)^{-\frac{1}{2}}$. The nine point implicit difference schemes (\ref{eqn: 2layer scheme a}) and (\ref{eqn: 2layer scheme b}), with tri-diagonal matrices of constant coefficients, are solved simultaneously using a Thomas Algorithm (e.g., \cite{thomas}).  

A single Boussinesq equation with arbitrary coefficients (i.e. system (\ref{fg}) for $\delta=\gamma=0$ and $f = 0$) is used to examine the scheme's stability. This solution can be approximated by the solution of the difference scheme (\ref{eqn: 2layer scheme b}) with $\gamma=0$. We first linearise this scheme by setting $g_{i,j}=g_0+\tilde{g}_{i,j}$ where $g_0$ is a constant such that $g_0 \ge g_{i,j} \ \forall i,j$. Using a Von-Neumann stability analysis we substitute $g_{i,j}=G^je^{\text{i}\theta ih}$ (where $\text{i}^2 = -1$) into the linearised version of (\ref{eqn: 2layer scheme b}) with $\gamma=0$ and derive 
\begin{eqnarray}
G^2-2\mu G+1=0\ \ \ {\rm{where}} \ \ \  \mu=1-\frac{2k^2 (c^2 + \alpha g_0) {\rm{sin}}^2 \frac{\theta h}{2}}{h^2 +4 \beta {\rm{sin}}^2 \frac{\theta h}{2}}.
\label{eqn: G}
\end{eqnarray}

\noindent For stability we require $|G|\leq1$ $\forall\theta$ and arbitrary $k,h$, which is true provided $|\mu|<1$ and thus implies $k<k_c=\sqrt{\frac{h^2+4\beta}{c^2+\alpha g_0}}$. Hence the roots of the quadratic in $G$ have modulus one and the linearised form of  the difference sheme (\ref{eqn: 2layer scheme b}) with $\gamma=0$ is stable provided $k<k_c$. (In practice, we used a stricter condition $k < \frac 12 k_c$, to accommodate for the effects of nonlinearity). It can be shown that the principal truncation error of this scheme is $O(h^2k^2)$.

Numerical simulations for the symmetric case (scheme (\ref{eqn: 2layer scheme a}) with $\delta=0$) compared with the known analytical solution (\ref{soliton}) reveal that the choice of discretisation can reduce the maximum error, across $x$ for a given time, to as low as $O(10^{-5})$. This accuracy is within the range which is deemed suitable from previous work on Boussinesq-type equations (see Refs. \cite{P8}-\cite{P4}). The step size $h=k=0.01$ results in errors of this order and is thus chosen for our simulations. We also numerically approximate $u(x,t)$ and $w(x,t)$ via Simpson's rule using the relationship ${\int^x_{-L}f,g\ dx=u,w}$, due to the boundary conditions (\ref{eqn: 2layer BC}), and hence utilise the energy conservation law given in Section 1, although integration introduces additional errors. The conserved quantity $\int_{-\infty}^{\infty} A_2 dx$ was monitored and for simulations with $h=k=0.1$  the energy was conserved,  within the chosen time interval, up to 0.021\% and 0.006\% for the results shown in Fig.2 and Fig.3, respectively. For smaller step sizes these computations become very time-consuming, but there are no noticeable differences in the plots of the solutions for $h=k=0.1$ and $h=k=0.01$. (A useful discussion of the difficulties associated with the accuracy of conservation laws in finite-difference schemes can be found in Ref. \cite{C1}.)

\begin{figure}[h]
\begin{center}
\includegraphics[width=4.5cm]{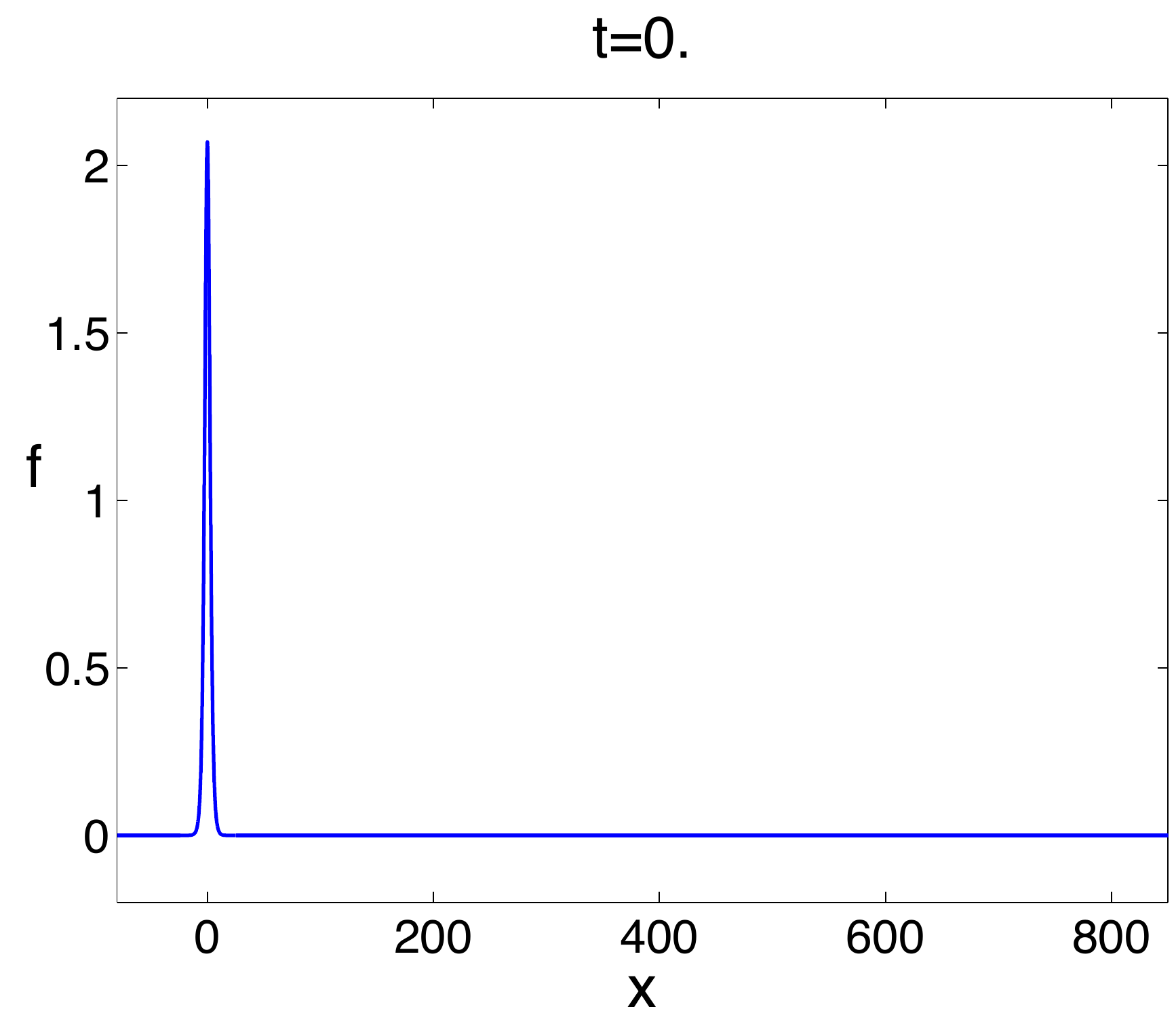} \quad 
\includegraphics[width=4.5cm]{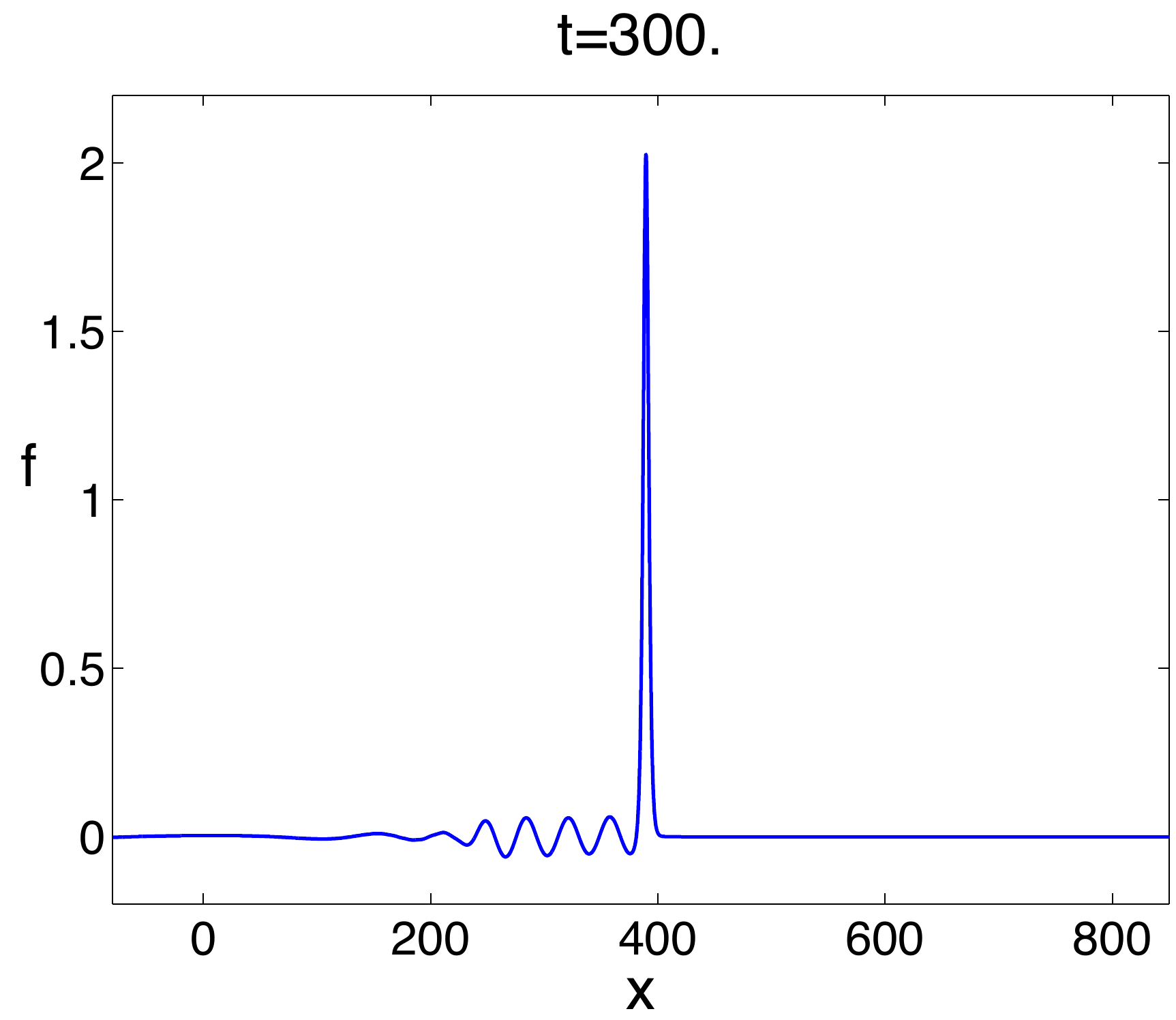} \quad 
\includegraphics[width=4.5cm]{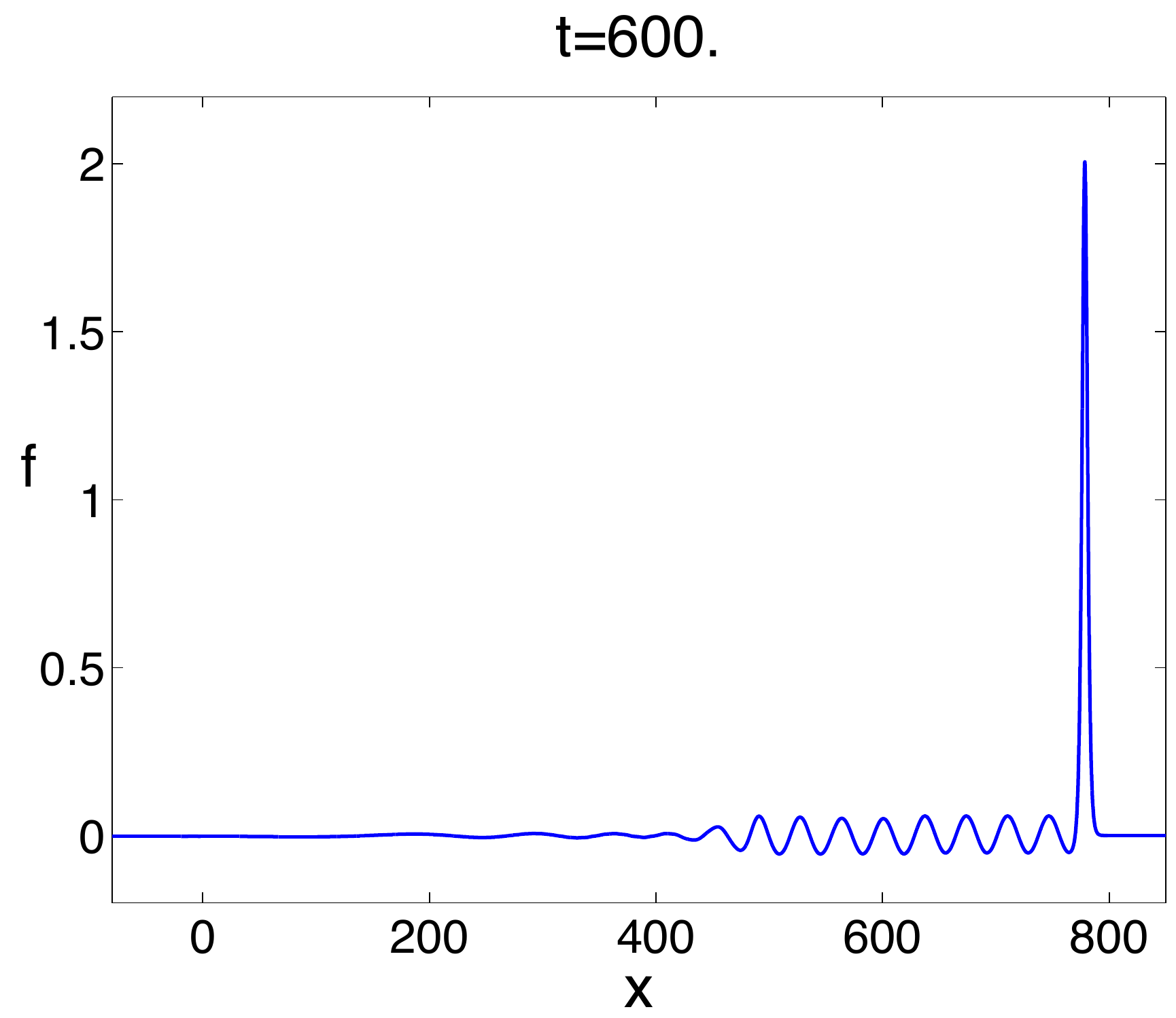} \quad \\
\includegraphics[width=4.5cm]{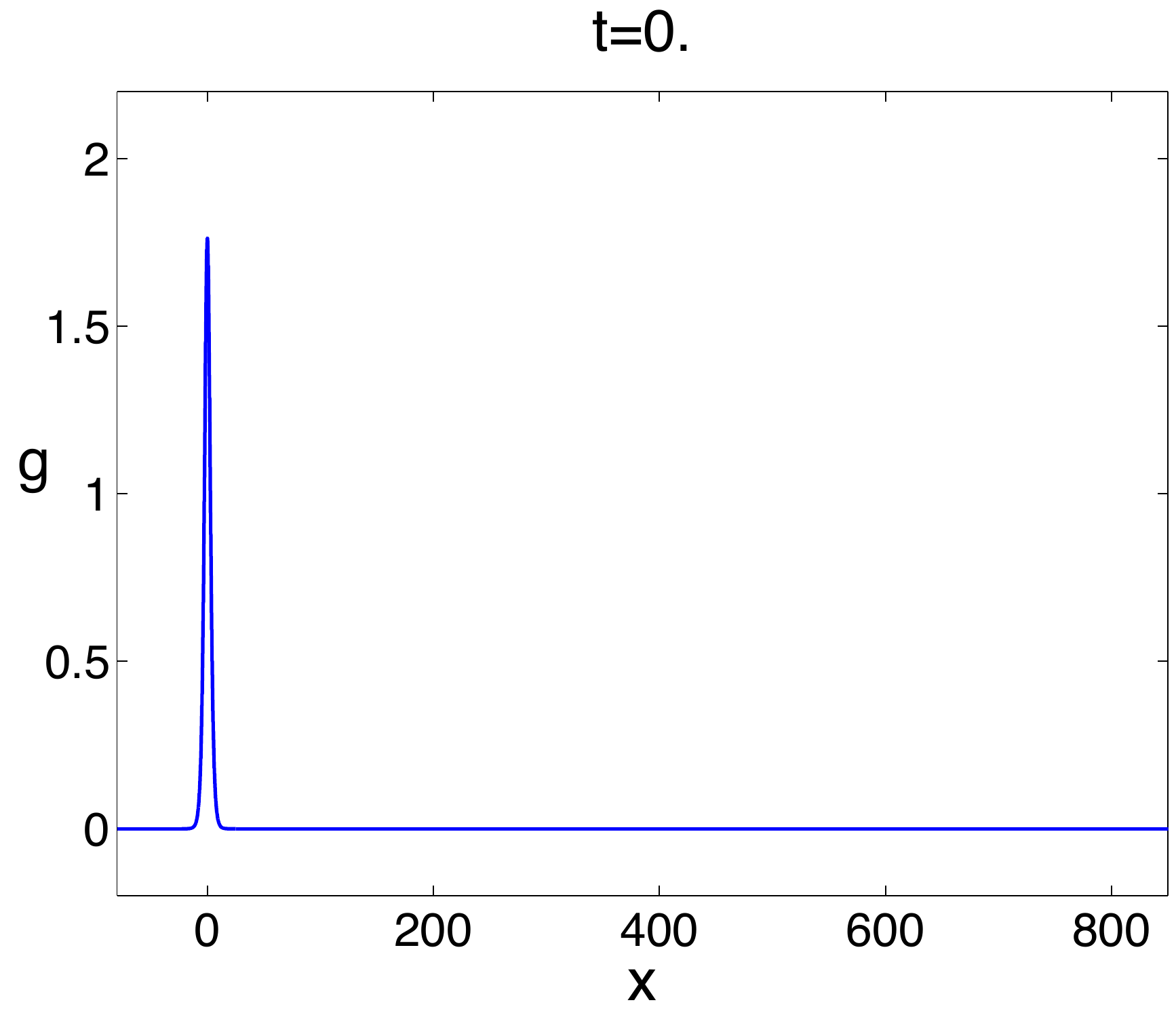} \quad 
\includegraphics[width=4.5cm]{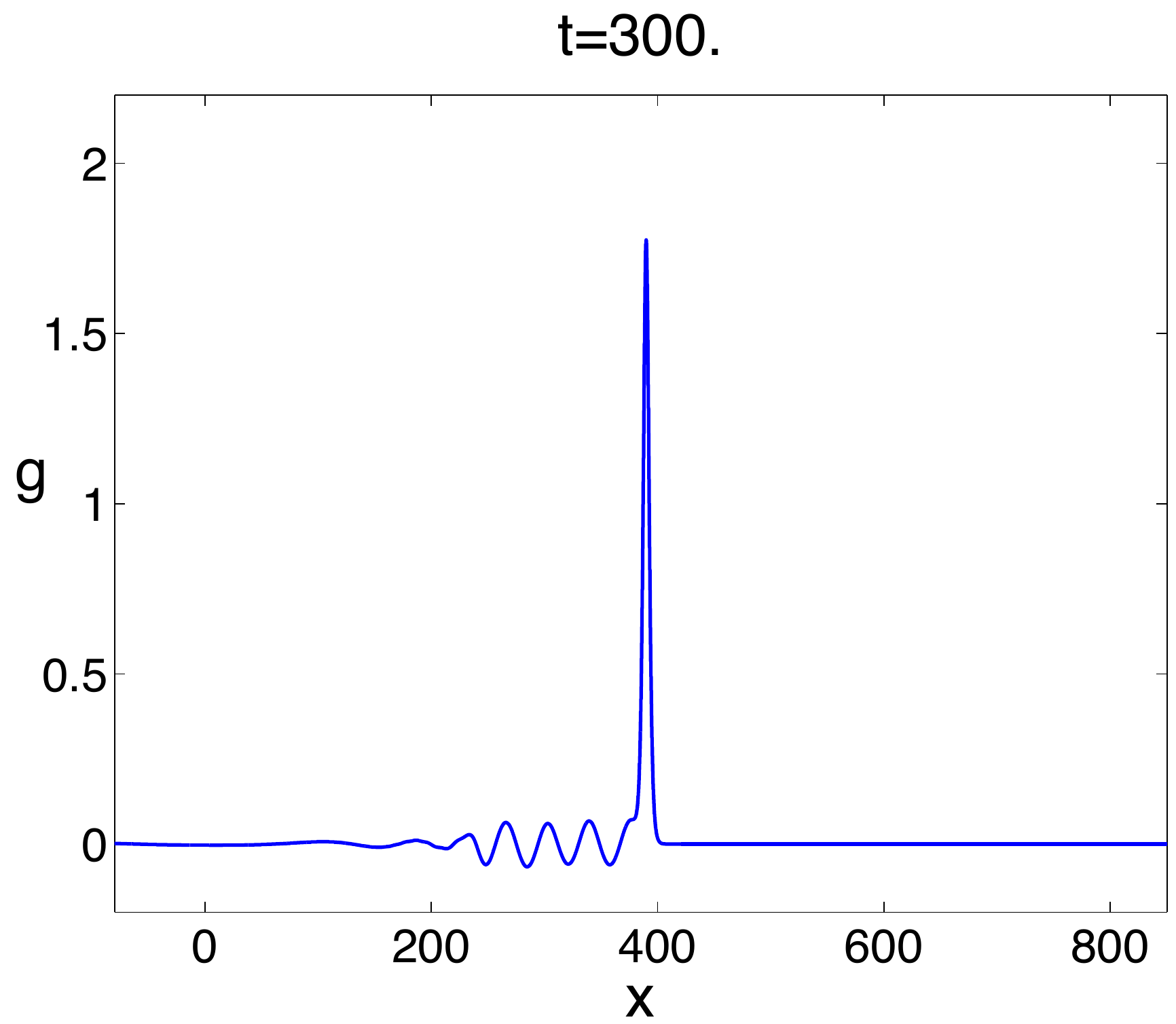} \quad 
\includegraphics[width=4.5cm]{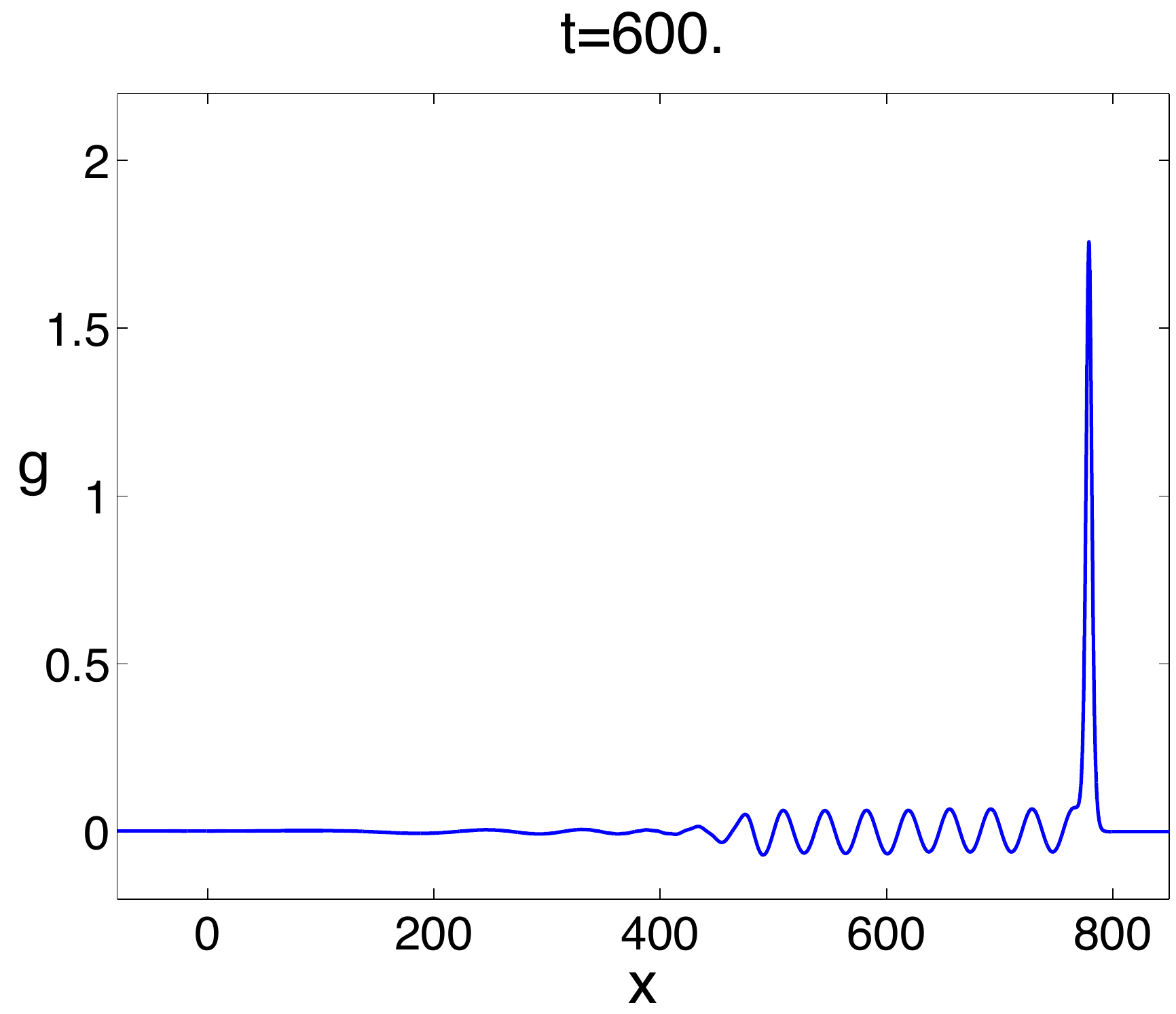} 
\caption{\small Generation of a radiating solitary wave for $c=1.05, \alpha = \beta = 1, \gamma = \delta = 0.01; v_1=v_2 = 1.3$ from  pure solitary waves of the uncoupled equations.}
\label{eps}
\end{center}
\end{figure}

\begin{figure}[h]
\begin{center}
\includegraphics[width=4.5cm]{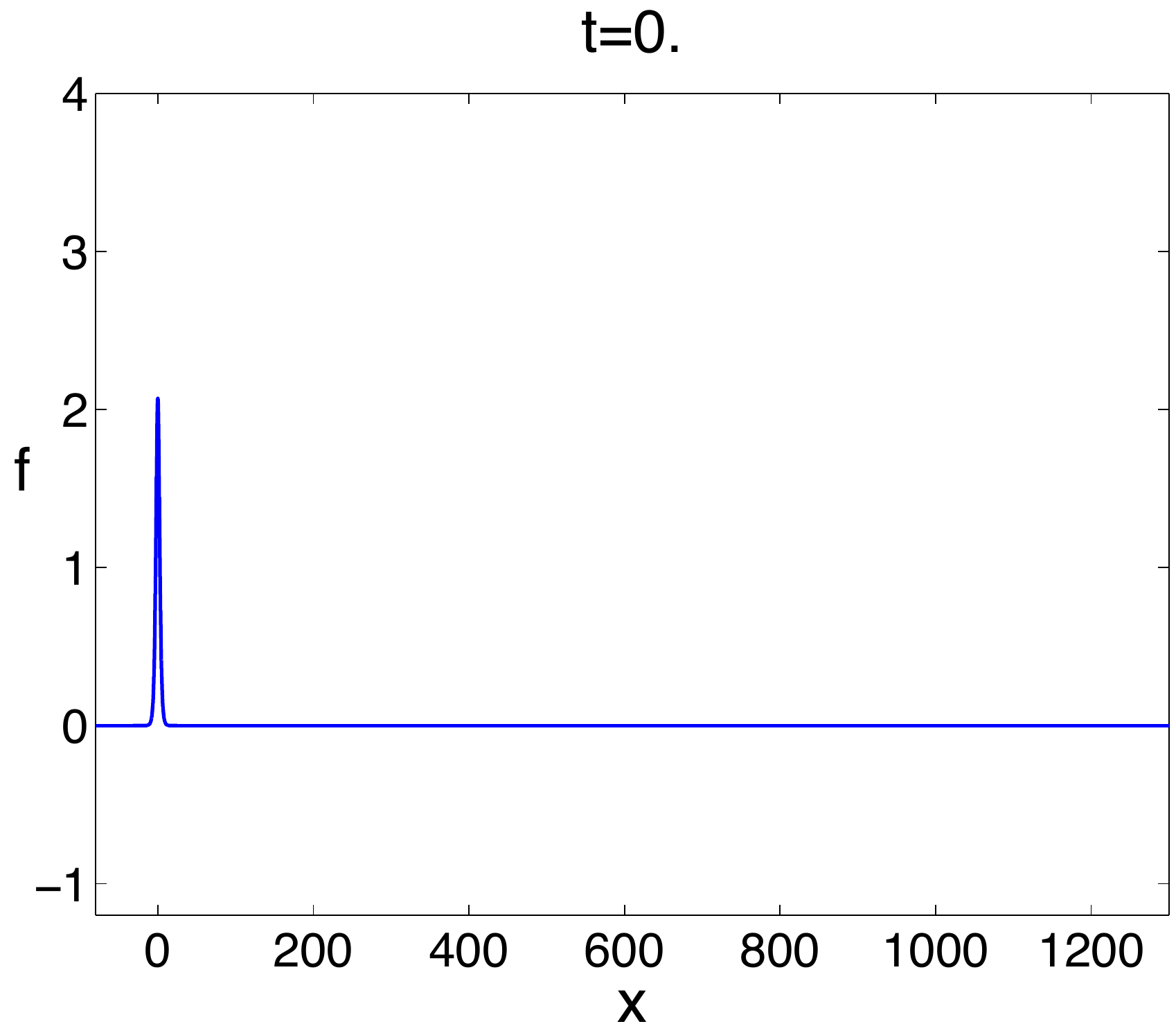} \quad 
\includegraphics[width=4.5cm]{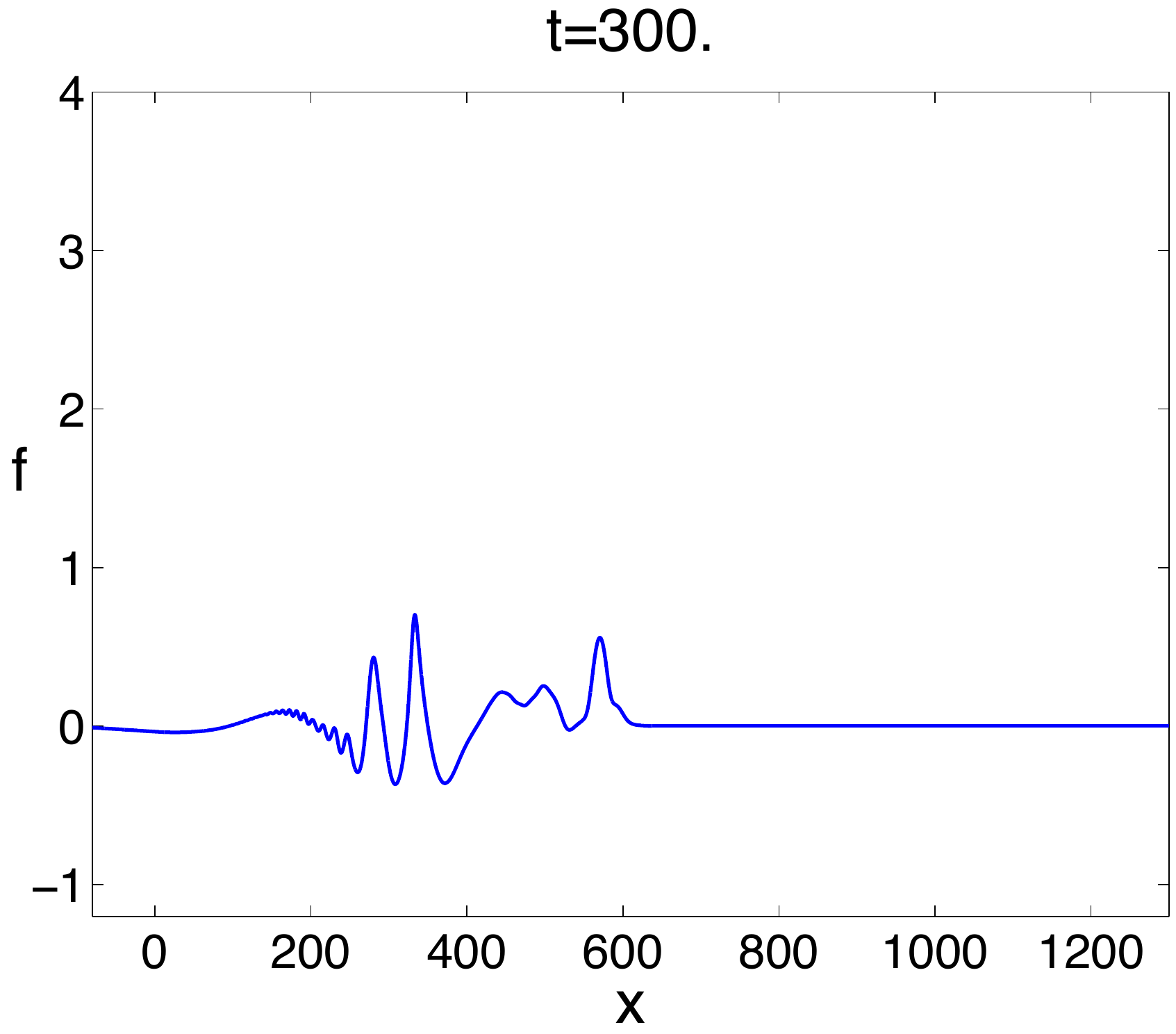} \quad 
\includegraphics[width=4.5cm]{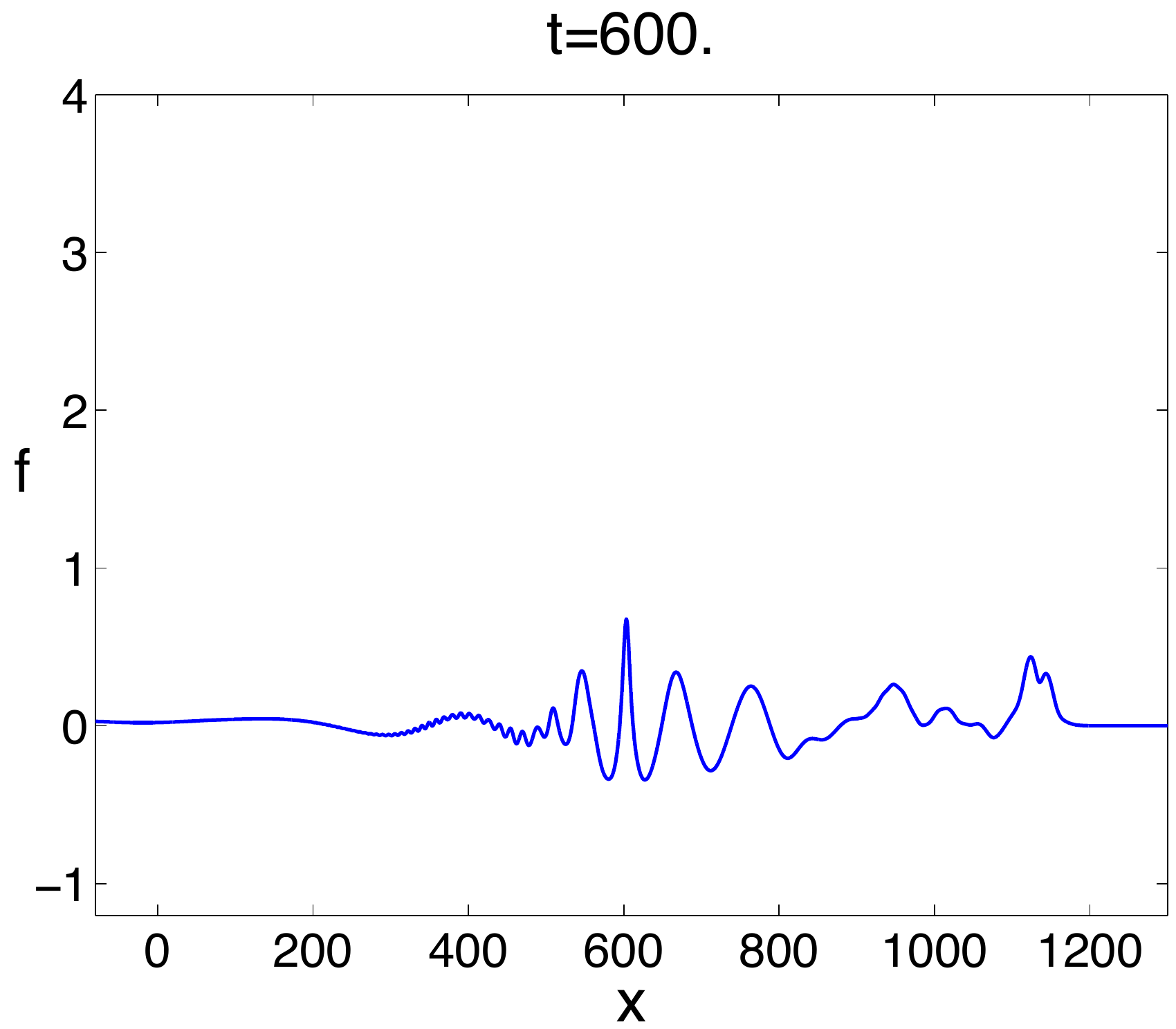} \quad \\
\includegraphics[width=4.5cm]{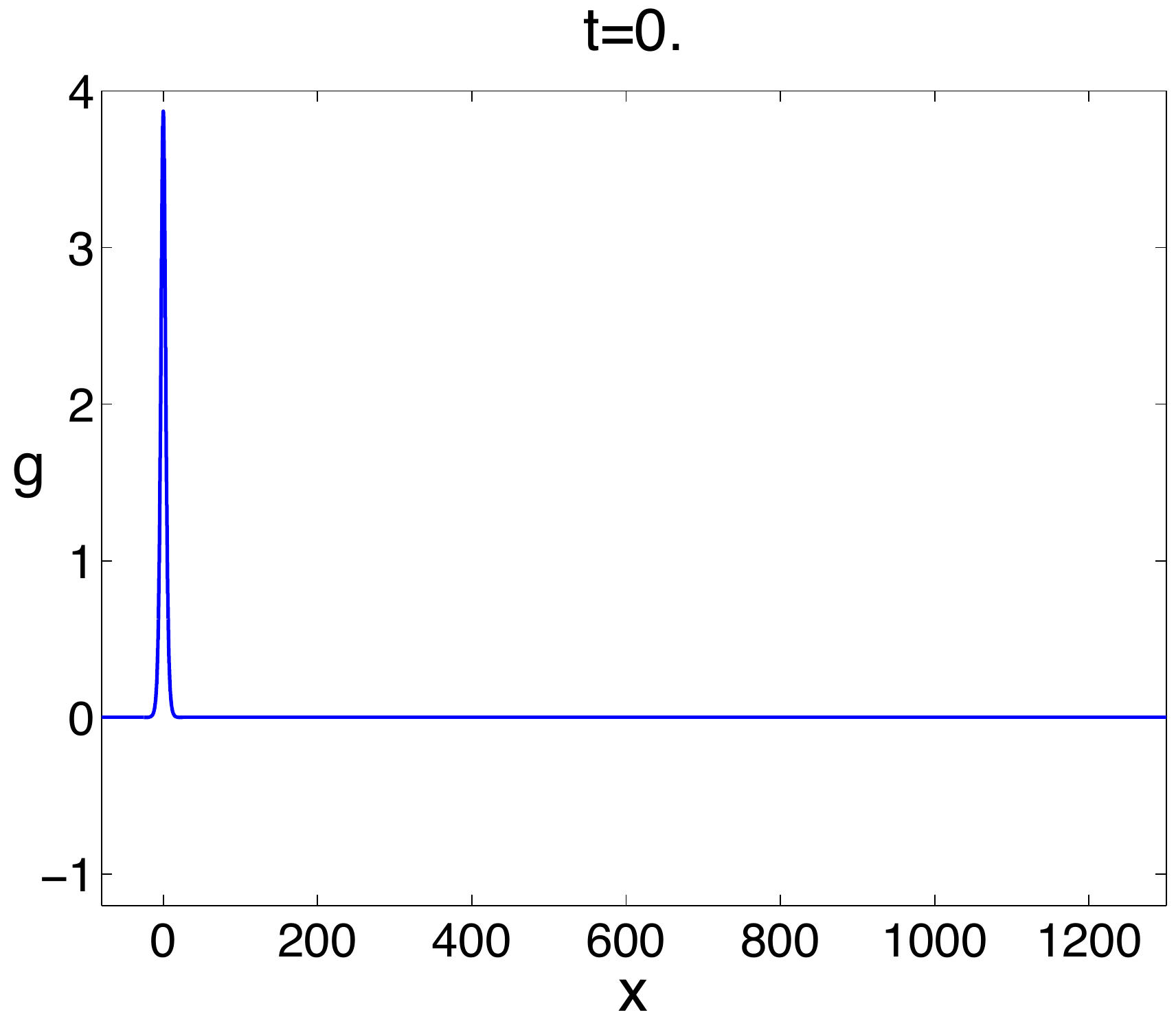} \quad 
\includegraphics[width=4.5cm]{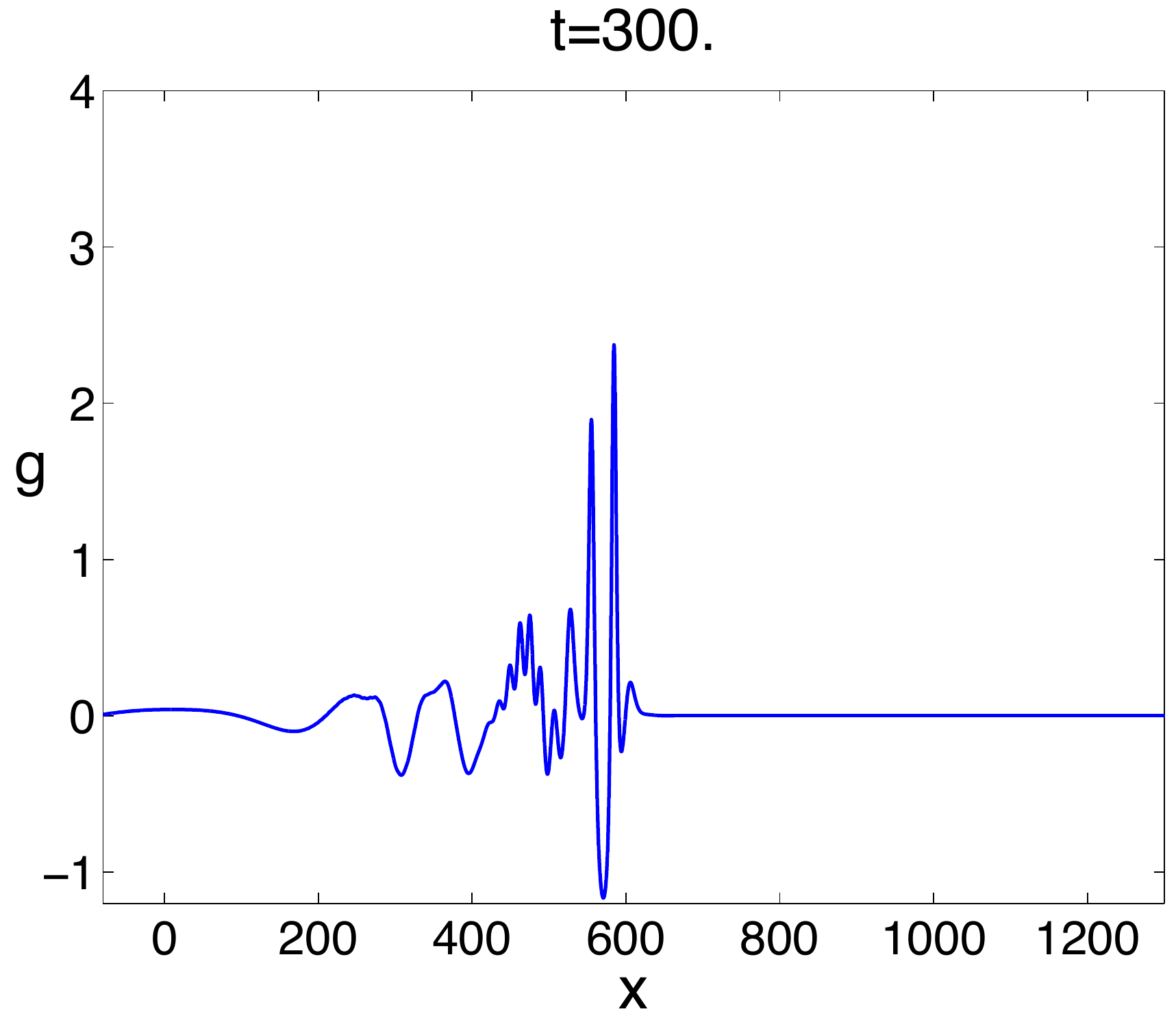} \quad 
\includegraphics[width=4.5cm]{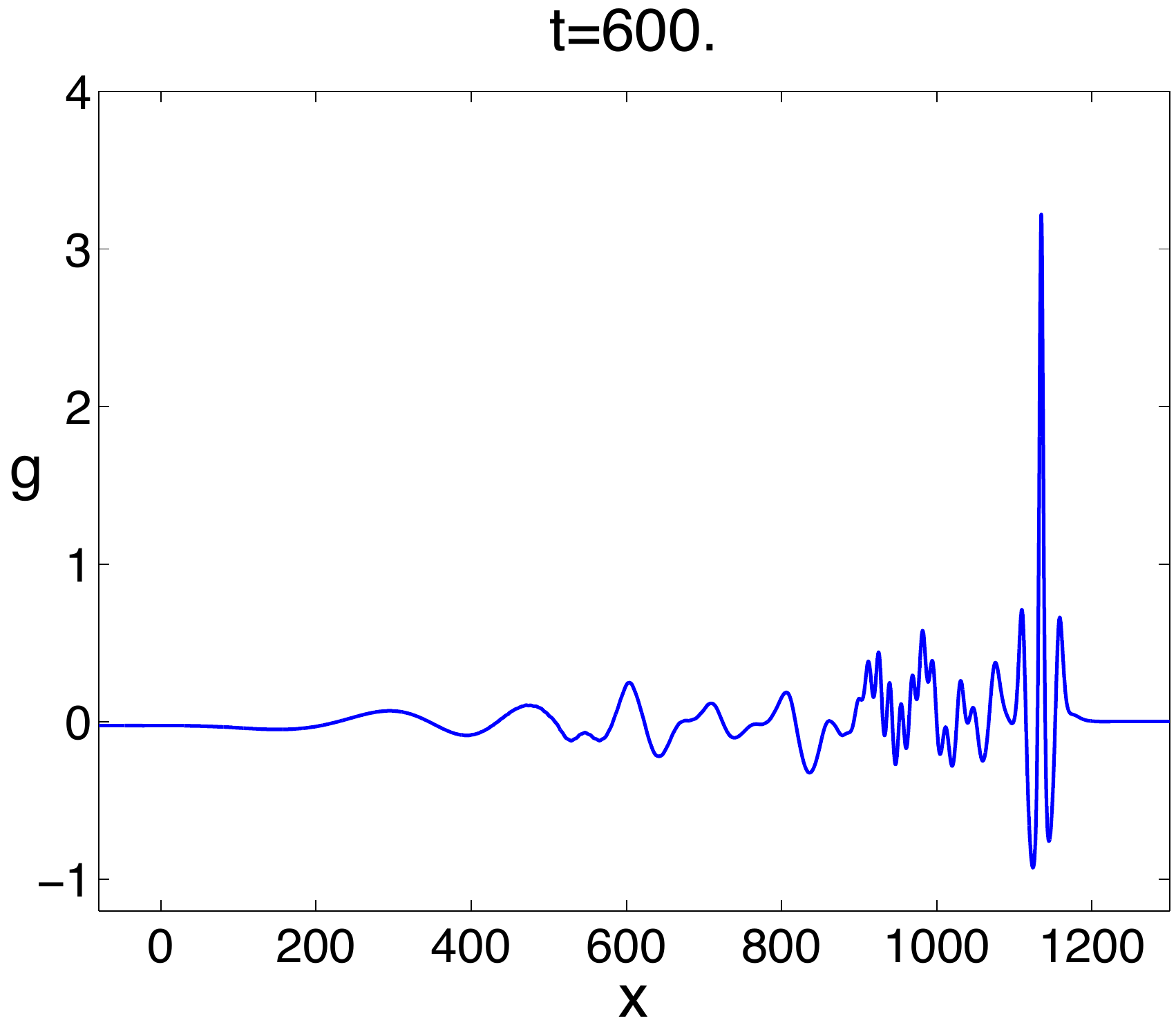} 
\caption{\small Generation of wave packets for $c=2, \alpha = \beta = 1, \gamma = \delta = 0.01; v_1=1.3, v_2 = 2.3 $ from pure solitary waves of the uncoupled equations.}
\label{one}
\end{center}
\end{figure}
%\newpage

The emergence of radiating solitary waves replacing the initial pure solitary waves in both components of $f$ and $g$ is shown in Fig.\ref{eps} for the case $c=1.05, \beta=\alpha=1, \delta=\gamma=0.01; v_1=v_2=1.3$. This result agrees with the discussion in Section 2 and numerical studies in Ref. \cite{KSZ}. The numerically determined wavelength of the oscillatory tail at $t=300$ is $(36.5 \pm 0.1)$ for $f$ and $(37.0 \pm 0.2)$ for $g$, which is close to the theoretical prediction $(\approx 36.7)$ for $p=1.3$ using the dispersion relation (see Fig.1).

For the case $c=2, \beta=\alpha=1, \delta=\gamma=0.01; v_1=1.3, v_2=2.3$, the initial solitary waves are replaced by dominant wave packets in both components, shown in Fig.\ref{one}. The emergence of a wave packet in both $f$ and $g$ can be observed almost instantaneously and for long time appears stable. From our asymptotic analysis for $c-1=O(1)$ in Section 3 we found that to leading order the solution for both $f$ and $g$, for right propagating waves, are the solution to the two Ostrovsky equations (\ref {eqn: one(f-) }) and (\ref {eqn: one(g-) }). At $t=600$ in Fig.\ref{one} the leading wave packet  in $g$ is qualitatively very similar to the numerical solution of the Ostrovsky equation studied in Ref. \cite{GH}. Similarly, at $t=600$ for $f$ a similar but smaller and slower moving wave packet  is present around $x = 600$. In Ref. \cite{GH} it is shown that a parameter $a_0$, equivalent to $a_{f0}=\frac{6(v_1-1)}{\sqrt{\delta}}$ and $a_{g0}=\frac{6(v_2-c)}{\sqrt{\beta\delta}}$ for $f$ and $g$ respectively in our system, determines the emergence of this distinct wave packet. For our simulations in Fig.\ref{one}, this parameter lies within the range for which the distinctive wave packet will emerge and hence it can be seen in both $f$ and $g$. As $a_{f0},a_{g0}\rightarrow 0$ there exists a range for which there is no emergence of a distinctive wave packet. Alternatively as the parameters are increased, the faster the wave packet emerges and in the case of $g$, the faster it will move away from the rest of the solution. The range for our results for which this transition occurs is also in very good agreement with the results for the Ostrovsky equation in Ref. \cite{GH}.

The simulations in this section confirm that there is a distinctive difference in the qualitative behaviour of the solution upon varying the difference in the characteristic speeds,  and hence support our asymptotic analysis. From our numerics we can conclude that for pure solitary wave initial data in the parameter range $c-1=O(\epsilon)$, stable radiating solitary waves emerge in both components $f$ and $g$. However as we increase the difference in the characteristic speeds, i.e. let $c-1=O(1)$, we see the emergence of wave packets, agreeing with the numerical solution of the Ostrovsky equation \cite{GH}, in both components $f$ and $g$.

%%%%%%%%%%%%%%%%%%%%%%%%%%%%% CONCLUSION %%%%%%%%%%%%%%%%%%%%%%%%

\section{Conclusions}

In this paper we addressed the question of constructing a weakly nonlinear solution of the initial-value problem for coupled Boussinesq-type equations for localised or sufficiently rapidly decaying initial data, generating sufficiently rapidly decaying right- and left-propaga\-ting waves. Crucially, we considered the general case, when the two linear wave operators have different characteristic speeds, which complicates the analysis since in this case the number of characteristic variables (four) is greater than the number of independent variables (two). Further generalisations to the case of more than two equations (and characteristic speeds) are straightforward. 

We introduced two different types of asymptotic multiple-scales expansions for the cases $c-1 = O(\epsilon)$ and $c-1 = O(1)$ and averaged with respect to the fast time, which allowed us to derive to leading order a hierarchy of asymptotically exact coupled and uncoupled Ostrovsky equations for unidirectional waves. We then constructed the nonsecular solution of the initial-value problem in terms of solutions of the derived leading order models for the values of time up to $O(\epsilon^{-1})$,  within the accuracy of the problem formulation.  To construct a more accurate solution, and for greater values of time,  one needs to know higher-order terms in the original cRB Eqs. (\ref{fg1}).

We performed numerical simulations of the original unapproximated coupled regularised Boussinesq Eqs. (\ref{fg1}) for the initial conditions in the form of co-propagating pure solitary waves of the uncoupled equations, and compared our numerical results with the known numerical results for the Ostrovsky equation \cite{GH}, which confirmed predictions of our leading order asymptotic theory. Expanded numerical studies for this and other types of initial conditions are currently underway. 

The approach developed in this paper is generic and can be used to construct weakly nonlinear solutions of some other initial-value problems, and in other physical contexts. In particular, it is interesting to derive a hierarchy of Ostrovsky equations and consider the initial-value problem for the original area of appearance of this equation (rotating ocean), which will be discussed somewhere else. 

When $\delta = \gamma = 0$, our solution yields 
an explicit weakly nonlinear solution of the initial-value problem for  the single Boussinesq equation 
$$
f_{tt} - f_{xx} = \epsilon \left [\frac 12 (f^2)_{xx} + f_{ttxx}\right ], \quad f|_{t=0} = F(x), \quad f_t|_{t=0} = V(x),
$$
(or any other asymptotically equivalent form of this equation) for the case when the initial conditions generate sufficiently rapidly decaying right- and left-propagating waves (i.e. $\int_{-\infty}^{\infty} V(x) dx = 0$).
 The solution has  the following form:
$$
f = f^-(\xi, T) + f^+(\eta, T) + \epsilon \left  [-\frac 14 \left  ( 2  f^- f^+ + f^-_{\xi} \int f^+ d \eta + f^+_{\eta} \int f^- d \xi \right )  + \phi(\xi, T) + \psi(\eta, T)\right ] + O(\epsilon^2),
$$
where $\xi = x-t, \eta = x+t, T = \epsilon t$. The functions $f^-$ and  $f^+$ are
solutions of the  initial-value problems for two Korteweg-de Vries equations
\begin{eqnarray*}
&&f^-_T + \frac 12 f^- f^-_\xi + \frac 12 f^-_{\xi \xi \xi} = 0, \quad f^+_T - \frac 12 f^+ f^+_\eta - \frac 12 f^+_{\eta \eta \eta} = 0,\\
&&f^{\pm}|_{T=0} = \frac 12 \left [F(x \pm t) \pm \int_{-\infty}^{x \pm t} V(x) dx\right ],
\end{eqnarray*}
integrable by the Inverse Scattering Transform \cite{GGKM} (see also Refs.  \cite{AS,DJ} and  the relevant discussion of matching of the {\it near-field} and {\it far-field} solutions for unidirectional waves in Ref. \cite{J}). The functions $\phi$ and $\psi$ are given by the formulae
\begin{eqnarray*}
\phi(\xi, T) = \frac 12 \left [P(\xi, T) + \int_{-\infty}^\xi Q(x,T) dx\right ], \quad
\psi(\eta, T) = \frac 12 \left [P(\eta, T) - \int_{-\infty}^\eta Q(x,T) dx \right ], 
\end{eqnarray*}
where
\begin{eqnarray*}
&&P(x, T) =  \frac 14 \left  [2  f^- f^+ + f^-_{\xi} \int f^+ d \eta + f^+_{\eta} \int f^- d \xi \right ]_{t=0}, \\
&& Q(x, T) =\left [f^-_T + f^+_T + \frac 14 \left (f^+ f^-_\xi  - f^- f^+_\eta + f^-_{\xi \xi} \int f^+ d \eta - f^+_{\eta \eta} \int f^- d \xi   \right )\right ]_{t=0}
\end{eqnarray*}
(within the accuracy $O(\epsilon^2)$ of the problem formulation, the dependence of the functions $\phi$ and $\psi$ on $T$ is inherited from the dependence of the leading order functions $f^-$ and $f^+$, or it may be neglected, at least for sufficiently small values of time).

For the practical applications of the constructed solution 
%for the cRB system (\ref{fg1}) 
it is useful to remember that within the accuracy of the problem formulation in (\ref{fg1}) (i.e. $O(\epsilon^2)$), the initial conditions (\ref{IC1}) and  (\ref{IC2}) can be represented in the form
\begin{eqnarray*}
&&f|_{t=0} = F^0(x) + \epsilon F^1(x) + O(\epsilon^2), \quad g|_{t=0} = G^0(x) + \epsilon G^1(x) + O(\epsilon^2), \\
&&f_t|_{t=0} = V^0(x) + \epsilon V^1(x) + O(\epsilon^2), \quad g_t|_{t=0} = W^0(x) + \epsilon W^1(x) + O(\epsilon^2), 
\end{eqnarray*}
which not only allows one to formally satisfy the zero mass constraints for $f_0^{\pm}, g_0^{\pm}$ by adding appropriate $O(\epsilon^2)$ `pedestal'  terms, as explained in section 3, but also gives us some flexibility with the choice of initial conditions for the auxiliary IVP problems for unidirectional waves by splitting the functions $F(x), G(x), V(x), W(x)$ into  a `nice' $O(1)$ part (i.e. such that the IVP problems have some favourable analytical properties, e.g. from the viewpoint of the Inverse Scattering Transform when these are for the KdV equations) and a small $O(\epsilon)$ remainder, which can be readily accounted for in D'Alembert's-like formulae (\ref{phipsi}) and (\ref{phipsinew}) for the functions $\phi_i$ and $\psi_i, i=1,2.$

Finally, we would like to emphasise the importance of the Ostrovsky equation as a canonical asymptotically exact model, similar to the Korteweg-de Vries model. The reduced form of the Ostrovsky equation 
$$
(\eta_t + \nu \eta \eta_x)_x = \lambda \eta
$$
was recently shown to be an integrable equation \cite{VP1,Kraenkel}, reducible to the Tzitzeica equation \cite{Tzi}. We believe that the full Ostrovsky equation also might have some `nice' analytical properties (although it is not necessarily integrable in the conventional  sense).

\section{Acknowledgments}

We thank R.H.J. Grimshaw and L.A. Ostrovsky for references and useful discussions about the Ostrovsky equation, M.J. Ablowitz  for the discussion of the initial time layer in Ref. \cite{AW}, E.V. Ferapontov for the reference \cite{Kraenkel}, and C. Klein for the helpful advice on numerical simulations.

 %\pagebreak


\begin{thebibliography}{99}

\bibitem{Ostrovsky} L.A. Ostrovsky, Nonlinear internal waves in a rotating ocean, Oceanology 18
%(2) 
(1978) 119-125.

\bibitem{Leonov} A.I. Leonov, The effect of the earth's rotation on the propagation of weak nonlinear surface and internal long oceanic waves, Ann. NY Acad. Sci. 373 (1981) 150-159.

\bibitem{GHO} R.H.J. Grimshaw, J.-M. He, L.A. Ostrovsky, Terminal damping of a solitary wave due to radiation in rotational systems, Stud. Appl. Math. 101 (1998) 197-210.

\bibitem{Helfrich} K.R. Helfrich, Decay and return of internal solitary waves with rotation, Phys. Fluids 19  (2007) 026601.

\bibitem{GH} R. Grimshaw, K. Helfrich, Long-time solutions of the Ostrovsky Equation, Stud. Appl. Math. 121 (2008)  71-88.

\bibitem{YK} D. Yagi, T. Kawahara, Strongly nonlinear envelope soliton in a lattice model for periodic structure, Wave Motion 34 (2001) 97-107.

\bibitem{Gerkema} T. Gerkema, A unified model for the generation and fission of internal tides in a rotating ocean, J. Mar. Res. 54 (1996) 421-450.
%421-450

\bibitem{KSZ} K.R. Khusnutdinova, A.M. Samsonov, A.S. Zakharov, Nonlinear layered lattice model and generalized solitary waves in layered elastic structures, Phys. Rev. E 79  (2009) 056606.


\bibitem{BBM} T.B. Benjamin, J.L. Bona, J.J. Mahony, Model equations for long waves in nonlinear dispersive systems, Philos. Trans. R. Soc. Lond. A 272 (1972) 47-48.

\bibitem{Christov} M.A. Christou, C.I. Christov, Interacting localized waves for the regularized long wave equation via a Galerkin spectral method, Math. Comp. Sim. 69 (2005) 257-268.


\bibitem{Z} V.E. Zakharov,
On stochastisation of one-dimensional chains of nonlinear oscillators, 
%{\it Sov. Phys. JETP } %{\bf 65} (1973) 219-225.
Sov. Phys. JETP 38 (1974) 108-110.
%(1974) 108-110.


\bibitem{AS} M.J. Ablowitz, H. Segur, Solitons and the Inverse Scattering Transform, SIAM Philadelphia (1981).

\bibitem{Samsonov1} A.M. Samsonov,
Soliton evolution in a rod with variable cross section,
%{\it Sov. Physics - Doklady } {\bf 29} (1984) 586-588.
Sov. Phys. Dokl. 29 (1984) 586-588.

\bibitem{PS} A.V. Porubov and A.M. Samsonov, 
Refinement of longitudinal strain wave propagation in non-linearly elastic rod, 
Sov. Technic. Phys. Lett. 19
%12, 365-366
(1993) 365-366.

\bibitem{Samsonov2}  A.M. Samsonov, Strain Solitons in Solids and How to
Construct Them, Chapman and Hall/CRC, Boca Raton, 2001.

\bibitem{Porubov} A.V. Porubov, Amplification of Nonlinear Strain Waves in Solids, World Scientific, Singapore, 2003.

\bibitem{JE} J. Janno and J. Engelbrecht, Solitary waves in nonlinear microstructured materials, J. Phys. A: Math. Gen. 38 (2005) 5159-5172.
%5159-5172.

\bibitem{Maugin} G.A. Maugin, Nonlinear Waves in Elastic Crystals, Oxford University Press, Oxford, 1999.

\bibitem{C1} C.I. Christov and G.A. Maugin, An implicit difference scheme for the long-time evolution of localized solutions of a generalized  Boussinesq system, J. Comp. Phys. 116 (1995) 39-51.

\bibitem{C2} C.I. Christov, G.A. Maugin, M.G. Velarde, Well-posed Boussinesq paradigm with purely spatial higher-order derivatives, Phys. Rev. E 54 (1996) 3621-3638.

\bibitem{C3} C.I. Christov, T.T. Marinov, R.S. Marinova, Identification of solitary-wave solutions as an inverse problem: Application to shapes with oscillatory tails, Math. Comp. Sim. 80 (2009) 56-65.

\bibitem{FPU} E. Fermi, J. Pasta, S. Ulam, Studies on nonlinear problems, I, Los Alamos Scientific Laboratory Report No. LA-1940 (1955). Reprinted in A.C. Newell (Ed.), Nonlinear Wave Motion, AMS Lect. Appl. Math. 15 (1974) 143-156.

\bibitem{Miles1} J.W. Miles, Obliquely interacting solitary waves, J. Fluid Mech. 79 (1977) 157-169.

\bibitem{Miles2} J.W. Miles, Resonantly interacting solitary waves, J. Fluid Mech. 79 (1977) 171-179.

\bibitem{GG} J.A. Gear and R. Grimshaw, Weak and strong interactions between internal solitary waves, Stud. Appl. Math. 70 (1984) 235-258.

\bibitem{Fission} K.R. Khusnutdinova, A.M. Samsonov, Fission of a longitudinal strain solitary wave in a delaminated bar, Phys. Rev. E 77  (2008) 066603.


\bibitem{BGK} E.S. Benilov, R. Grimshaw, E.P. Kuznetsova, The generation of radiating waves in a singularly perturbed Korteweg-de Vries equation, Phys. D 69 (1993) 270-278.


\bibitem{Shrira} V.V. Voronovich, I.A. Sazonov, and V.I. Shrira, On radiating solitons in a model of the internal wave-shear flow resonance, J. Fluid Mech. 568 (2006) 273-301.

\bibitem{Bona} J.L.  Bona, V.A. Dougalis and D.E. Mitsotakis, Numerical solution of Boussinesq systems of KdV-KdV type: II. Evolution of radiating solitary waves, Nonlinearity 21 (2008) 2825-2848.

%\bibitem{HVB} J.K. Hunter and J.-M. Vanden-Broeck,
%Solitary and periodic gravity-capillary waves of finite amplitude
%J. Fluid Mech. {\bf 134}, 205 (1982).
%205-219

\bibitem{VB} J.-M. Vanden-Broeck, Elevation solitary waves with surface tension, Phys. Fluids A 3 (1991)  2659-2663.
%2659-2663

%\bibitem{Beale} J.T. Beale,
%Exact solitary water waves with capillary ripples at infinity,
%Commun. Pure Appl. Math {\bf 44}, 211 (1992).
%211-247

%\bibitem{Sun} S.M. Sun,
%Existence of a generalized solitary wave with positive Bond number smaller than 1/3,
%J. Math. Anal. Appl. {\bf 156}, 471 (1991).
%471-504

\bibitem{Karpman} V.I. Karpman, Radiation by solitons due to higher-order dispersion, Phys. Rev. E 47 (1993) 2073-2082.

\bibitem{GJ} R. Grimshaw, N. Joshi, Weakly nonlocal solitary waves in a singularly perturbed Korteweg-de Vries equation,
SIAM J. Appl. Math. 55 (1995) 124-135.
%124-135


\bibitem{Boyd} J.P. Boyd, Weakly Nonlinear Solitary Waves and Beyond-All-Orders Asymptotics, Kluwer, Boston, 1998.

\bibitem{Lombardi} E. Lombardi, Oscillatory Integrals and Phenomena Beyond all Algebraic Orders, Lecture Notes in Mathematics 1741, Springer, Berlin, 2000.


\bibitem{GI} R. Grimshaw, G. Iooss, Solitary waves of a coupled Korteweg-de Vries system, Math. Comp. Sim. 62 (2003)  31-40.
%31-40

\bibitem{Grimshaw} C. Fochesato, F. Dias, R. Grimshaw, Generalized solitary waves and fronts in coupled Korteweg-de Vries systems, Physica D 210 (2005) 96-117.



\bibitem{Ch} A.R. Champneys, B.A. Malomed, J. Yang, D.J. Kaup, Embedded solitons: solitary waves in resonance with the linear spectrum, Phys. D 152-153 (2001) 340-354.
%340-354



\bibitem{Yang} J. Yang, Stable Embedded Solitons, Phys. Rev. Lett. 91 143903 (2003).

\bibitem{DSSK} G.V. Dreiden, A.M. Samsonov, I.V. Semenova, K.R. Khusnutdinova, Observation of a radiating bulk strain solitary wave in a solid waveguide, Techn. Phys. 81 (2011) 145-149.

\bibitem{GGK} S.D. Griffiths, R.H.J. Grimshaw, K.R. Khusnutdinova, Modulational instability of two pairs of counter-propagating waves and energy exchange in a two-component system, Phys. D 214 (2006) 1-24.

\bibitem{Duruk} N. Duruk, H.A. Erbay, A. Erkip, Blow-up and global existence for a general class of nonlocal nonlinear coupled wave equations, J. Diff. Eqs.  250 (2011) 1448-1459.

\bibitem{Duruk1} H.A. Erbay, private communication, 2011.


\bibitem{GM} R. Grimshaw and W.K. Melville, On the derivation of the modified Kadomtsev- Petviashvili equation, Stud. Appl. Math. 80 (1989)183Ð202.

\bibitem{AW} M.J. Ablowitz and X.-P. Wang, Initial time layers and Kadomtsev-Petviashvili-type	 equations, Stud. Appl. Math. 98 (1997) 121-137.

\bibitem{Horikis} T.P. Horikis, The short-pulse equation and associated constraints, J. Phys. A: Math. Theor. (2009) 442004. 

\bibitem{soerensen} M.P. Soerensen, P.L. Christiansen, P.S. Lomdahl, Solitary waves on nonlinear elastic rods.I, J. Acoust. Soc. Am. 76
%(3) 
(1984) 871-879.

\bibitem{thomas} W.F. Ames, Numerical Methods for Partial Differential Equations, Academic Press, Inc, Thomas Nelson and Sons Ltd., 1979.

\bibitem{P8} H.El-Zoheiry, Numerical study of the improved Boussinesq equation, Chaos Solitons Fractals 14 (2002) 377-384.

\bibitem{P9} A.G. Bratos, A predictor-corrector scheme for the improved Boussinesq equation, Chaos Solitons Fractals 40 (2009) 2083-2094.

\bibitem{P11} D. Irk, I. Dag, Numerical simulations of the improved Boussinesq equation, Numer. Meth. Partial Diff. Eqs. 26 (2009) 1316-1327.

\bibitem{P3} A. Mohsen, H. El-Zoheiry, L. Iskandar, A highly accurate finite-difference scheme for a Boussinesq-type equation, Appl. Math. Comp. 55 (1993) 201-212.

\bibitem{P6} M.A. Hajji, K. Al-Khaled, Analytic studies and numerical simulations of the generalized Boussinesq equation, Appl. Math. Comp. 191 (2007) 320-333.

\bibitem{P5} T. Ortega, J.M. Sanz-Serna, Nonlinear stability and convergence of the finite-difference methods for the ``good" Boussinesq equation, Numer. Math. 58 (1990) 215-229.

\bibitem{P4} H. El-Zoheriy, Numerical investigation for the solitary waves interaction of the ``good" Boussinesq equation, Appl. Numer. Math. 45 (2003) 161-173.

\bibitem{GGKM} C.S. Gardner, J.M. Greene, M.D. Kruskal, R.M. Miura, Method for solving the Korteweg-de Vries equation, Phys. Rev. Lett. 19 (1967) 1095-1097.

\bibitem{DJ} P.G. Drazin \& R.S. Johnson, Solitons: an introduction, Cambridge University Press, Cambridge, 1989.


\bibitem{J} R.S. Johnson, A Modern Introduction to the Mathematical Theory of Water Waves, Cambridge University Press, Cambridge, 1997.


\bibitem{VP1} V.O. Vakhnenko, E.J. Parkes, The two loop soliton solution of the Vakhnenko equation, Nonlinearity 11 (1998) 1457-1464.

%\bibitem{VP2} V.O. Vakhnenko, E.J. Parkes, The calculation of multi-soliton solutions of the Vakhnenko equation by the inverse %scattering method, Chaos, Solitons \& Fractals 13 (2002) 1819-1826.

\bibitem{Kraenkel} R. Kraenkel, H. Leblond, M.A. Manna, An integrable evolution equation for surface waves in deep water, Jan. 30 2011, arXiv: 1101.5773v1 [nlin.SI].

\bibitem{Tzi} M. Tzitzeica, Sur une nouvelle classe de surfaces, Comptes Rendus hebd. Seances l`Acad. Sciences Paris 150 (1910) 955, 1227.

%Tzitzeica, G. 1910. Sur une nouvelle classe de surfaces. Comptes Rendus, 150: 955Ð956?

\end{thebibliography}
\end{document}